\documentclass[a4paper,useAMS,usenatbib]{mnras}
\pdfoutput=1
\usepackage{newtxtext,newtxmath}
\usepackage[T1]{fontenc}
\usepackage{ae,aecompl}
\usepackage{graphicx}	
\usepackage{amsmath}	
\usepackage{listings}
\usepackage{dirtytalk}

\title[Mapping the tilt of the velocity ellipsoids]{Mapping the tilt of the Milky Way bulge velocity ellipsoids with ARGOS and $Gaia$ DR2}

\author[Iulia T. Simion et al]{Iulia T. Simion$^{1} \thanks{E-mail:isimion@shao.ac.cn}$, Juntai Shen$^{2, 1}$,  Sergey E. Koposov$^{3, 4, 5}$, Melissa Ness$^{6,7}$, 
\newauthor
 Kenneth Freeman$^{8}$, Jonathan Bland-Hawthorn$^{9}$ and Geraint F. Lewis$^{9}$\\
 $^{1}$Shanghai Astronomical Observatory, Chinese Academy of Sciences, 80 Nandan Road, Shanghai 200030, China \\
 $^{2}$Department of Astronomy, School of Physics and Astronomy, Shanghai Jiao Tong University, Shanghai 200240, China \\
  $^{3}$McWilliams Center for Cosmology, Department of Physics, Carnegie Mellon University, 5000 Forbes Avenue, Pittsburgh, PA, 15213, USA \\
  $^{4}$Institute of Astronomy, University of Cambridge, Madingley Road, Cambridge, CB3 0HA, UK \\
 $^{5}$Kavli
Institute for Cosmology, University of Cambridge, Madingley Road, Cambridge CB3 0HA, UK\\
$^{6}$Department of Astronomy, Columbia University, Pupin Physics Laboratories, New York, NY 10027, USA\\
$^{7}$Center for Computational Astrophysics, Flatiron Institute, 162 Fifth Avenue, New York, NY 10010, USA\\
$^{8}$Research School of Astronomy and Astrophysics, Australian National University, Cotter Rd., Weston, ACT 2611, Australia \\
$^{9}$Sydney Institute for Astronomy, School of Physics A28, University of Sydney, NSW 2006, Australia}
\date{Accepted XXX. Received YYY; in original form ZZZ}

\pubyear{2020}

\begin{document}
\label{firstpage}
\pagerange{\pageref{firstpage}--\pageref{lastpage}}
\maketitle

\begin{abstract}
Until the recent advent of $Gaia$ Data Release 2 (DR2) and deep multi-object spectroscopy, it has been difficult to obtain 6-D phase space information for large numbers of stars beyond 4 kpc, in particular towards the Galactic centre, where dust and crowding effects are significant. In this study we combine line-of-sight velocities from the Abundances and Radial velocity Galactic Origins Survey (ARGOS) spectroscopic survey with proper motions from {\it Gaia} DR2, to obtain a sample of $\sim$ 7,000 red clump stars with 3-D velocities.  We perform a large scale stellar kinematics study of the Milky Way (MW) bulge to characterize the bulge velocity ellipsoids. We measure the tilt $l_{v}$ of the major-axis of the velocity ellipsoid in the radial-longitudinal velocity plane in 20 fields across the bulge. The tilt or vertex deviation, is characteristic of non-axisymmetric systems and a significant tilt is a robust indicator of non-axisymmetry or bar presence. We compare the observations to the predicted kinematics of an N-body boxy-bulge model formed from dynamical instabilities. In the model, the $l_{v}$ values are strongly correlated with the angle ($\alpha$) between the bulge major-axis and the Sun-Galactic centre line-of-sight. We use a maximum likelihood method to obtain an independent measurement of $\alpha$, from bulge stellar kinematics alone. The most likely value of $\alpha$ given our model is $\alpha = (29 \pm 3)^{\circ}$. In the Baade's window, the metal-rich stars display a larger vertex deviation ($l_{v} = -40^{\circ}$) than the metal-poor stars ($l_{v} = 10^{\circ}$) but we do not detect significant $l_{v}-$metallicity trends in the other fields.
\end{abstract}




\section{Introduction}

Being the nearest bulge to us and therefore the most accessible for deep observations, the MW bulge has become the testbed for bulge formation theories in spiral galaxies. Over the past 20 years, instrumentation advances have allowed us to custom-build photometric and spectroscopic surveys (see \citealt{Babusiaux16review} for a surveys list and references therein) to answer important questions about the bulge origin, structure and evolution. Photometric surveys primarly focused on bright stars and were pivotal in revealing the bar morphology (\citealt{Stanek1994}, \citealt{Robin2012}, \citealt{Wegg2013}, \citealt{Simion2017}). Spectroscopic surveys were crucial in proving the dynamical origin of the bar by providing line-of-sight velocities (\citealt{Rich2007}, \citealt{Kunder2012}, \citealt{Ness2013kinematics}, \citealt{Nessapogee}).
Proper motions are difficult to measure at bulge distances of 4-12 kpc as they are intrinsically small and therefore require great accuracy. Initially, only a small number of $\sim$430 bulge stars possessed measurements of their transverse motions in a low extinction region named the Baade's window \citep{Spaenhauer1992}. This number increased by three orders of magnitude with the NASA/ESA Hubble Space Telescope (HST; \citealt{Rich2002, Koz2006, Clarkson2008, Soto2012, Soto2014}) and the Optical Gravitational Lensing Experiment II (OGLE II; \citealt{Sumi2004, Rattenbury2007}) which  provided proper motions with accuracies of the order of 0.9 - 3.5 mas/yr, particularly in low extinction fields or along the bulge minor axis. The new generation surveys, the Vista Variables in the Via Lactea survey \citep[VVV;][]{Minniti2010, Smith2018} and more recently $Gaia$ Data Release 2 ($Gaia$ DR2, \citealt{Prusti2016}), have released proper motions for tens and hundreds of millions of bulge stars with sub-milliarcsecond accuracy.

%
In this work we build a catalog of bulge stars with full phase-space information to study the bulge velocity ellipsoids. In particular, we search for evidence of bulge triaxiality in our sample which contains proper motions from $Gaia$ DR2 and radial velocities from the Abundances and Radial velocity Galactic Origins Survey \citep[ARGOS;][]{Freeman2013}. Stellar kinematics studies were late to show any evidence of bulge triaxiality compared to star counts, measurements of the integrated light and kinematics of the atomic and molecular gas studies, which were all providing strong evidence that the bulge is triaxial and rapidly rotating already by the early `90s \citep{Bulges1992}. The main difficulty was obtaining accurate measurements at bulge distances, especially in the highly dust-obscured regions. The first study of bulge triaxiality from kinematics used a sample of 62 K giants \citep{Zhao1994} with proper motions, radial velocities and metallicities in a low extinction bulge window, the Baade's window at $(l, b) = (1^{\circ}, -4^{\circ})$. The distributions of these stars projected onto three velocity planes ($v_{l}$-$v_{b}$, $v_{l}$-$v_{r}$ and $v_{r}$-$v_{b}$) were fitted by velocity ellipsoids with Gaussian profiles \citep[][]{Zhao1994}. Although the velocity distribution in the $v_{l}$-$v_{r}$ diagram was symmetric with respect to the $v_{r}$ and $v_{l}$ axes, the long axis of the velocity ellipsoid appeared tilted at an angle $l_{v}$ with the longitudinal velocity $v_{l}$ axis. The orientation of the axis of the velocity ellipsoid in the $v_{l}$-$v_{r}$ plane, $l_{v}$ or vertex deviation, is a measure of the correlation between the radial and longitudinal velocities and is affected by the bulge non-axisymmetry. In an axi-symmetric bulge, $l_{v}$ should be consistent with $l_{v} \sim 0^{\circ}$ along the minor axis ($l\sim0^{\circ}$). However, the metal-rich stars in Baade's window have $l_{v} \sim 40^{\circ}$ \citep{Zhao1994, Babusiaux2010, Soto2012}; this was the \say{first clear evidence for vertex deviation, a \say{smoking gun} of bulge triaxiality} \citep{Zhao1994}.  On the other hand, the  $v_{l}$-$v_{b}$ and $v_{r}$-$v_{b}$ diagrams did not display significant $l_{v}$.
\citet{Soto2007} confirmed this result with an expanded dataset of $\sim$300 stars, in the same region. The addition of spectroscopic measurements made it possible to study the variation of the vertex deviation with metallicity \citep{Babusiaux2010, Hill2011, Ness2013abundances},  suggesting that only the more metal rich stars display a tilted velocity ellipsoid distribution. For a review on the correlations between kinematics and metallicity prior to $Gaia$ DR2, see \citet{Babusiaux16review}.
Simulations have shown that the metal poor and metal rich components have different spatial distributions \citep{Debattista2017} which could explain the difference in the vertex deviation trends with metallicity. Perhaps the most complete 3D sample to date was provided by \citet{Soto2012}, who compiled a sample of $\sim$3200 stars observed by HST and VLT, in 6 bulge fields. They used HST proper motions and VLT/VIMOS Integral Field Unit (IFU) radial velocities with $\sim1$ mas/yr and 50 km/s accuracies respectively.

The sample we use in this work contains $\sim7000$ likely red clump (RC) bulge stars with $<$0.5 mas/yr proper motions and 1 km/s radial velocity accuracies respectively, distributed in 20 fields across the bulge, following the ARGOS footprint. RC stars are excellent standard candles, with a luminosity weakly dependent on age and metallicity, providing 5-10\% distance uncertainties (\citealt{Stanek1997}, \citealt{Girardi2016}, \citealt{Hawkins2017}). We could thus obtain the full phase-space information for our sample. The ARGOS fields of view are situated at latitudes beyond $4^{\circ}$ from the plane, avoiding the high extinction regions close to the Galactic plane, spiral arms and the long thin bar (e.g. \citealt{Wegg2015, Wegg2019}). Our catalog is suitable for studying the kinematics of the boxy/peanut bulge, successfully traced by star count studies using RC stars. Studies with RC stars have consistently reported that the bulge is triaxial with the major axis at an angle $\alpha \approx$ 20-30 degrees with respect to the Sun-Galactic Centre line (\citealt{Stanek1997}, \citealt{Wegg2013}, \citealt{Cao2013}, \citealt{Simion2017}). Asymmetries in the star counts (\citealt{Stanek1997}) show that the near end of the bar is situated at positive longitudes. 

This work investigates the relationship between the bulge velocity ellipsoid evidenced by our data sample and the bulge non-axisymetric density distribution induced by the viewing angle $\alpha$, with the help of a numerical model. Numerical models of a boxy bar/bulge where the angle $\alpha$ can be easily varied, are helpful to study the relationship between the two and interpret the observations.

Earlier bulge models \citep{Zhao1994, Hafner2000} were built and scaled to reproduce the morphological, chemical and kinematic properties of the MW, providing precious insight into the chemo-dynamical history of the bulge. It is generally agreed that the MW  hosts a boxy bulge (\citealt{Kormendy2004}), which forms from a bar instability in the disc and is subsequently thickened probably by the buckling instability \citep{Raha1991, Debattista2005, Martinez2006, Shen2010}. Other bar thickening mechanisms involving resonant heating were discussed by e.g. \citet{Combes1990}, \citet{Quillen2014} and  \citet{Sellwood2020}. The evolution of these bulges is affected by the exchange in angular momentum with the disc and dark halo, and the in-plane and vertical stellar motions (\citealt{Debattista2017}, \citealt{Fra2017}, \citealt{DiMatteo2019}). There is consensus between radial velocity \citep{Howard2009} and proper motion  \citep{Sanders19a, Clarke2019} surveys that the bulge rotates cylindrically with the rotational velocity profile almost independent of height, a behaviour that is well matched by a fully-evolutionary N-body model of a boxy/peanut bulge formed through the internal dynamical instabilities of the of the disc \citep{Shen2010, Qin2015}. Such a model also naturally explains the existence of an X-shaped structure visible at intermediate latitudes \citep{Nataf2010, McWilliam2010, Saito2011, Li2012, Nataf2014, Nataf2015, Shen2016, Ness2016wise}.  The  3D kinematics through the X-shape was studied by \citet{Vasquez2013}. While the Milky Way has an obvious boxy bulge, the presence of a `classical' bulge has not been completely excluded (\citealt{Shen2010, Saha2012}). `Classical' bulges form differently from boxy bulges, either through hierarchical merging \citep{Bender1992} or monolithic collapse \citep{Eggen1962}, in a similar fashion to mini-elliptical galaxies.

While our work falls in line with the studies of \citet{Zhao1994} and \citet{Soto2012}, there are several studies which use models to explain the observed  links between kinematics and metallicity, or morphology and metallicity. \citet{Athanassoula2017} found good qualitative agreement between the observed radial velocity dispersion variations in the bulge as a function of metallicity (\citealt{Ness2013kinematics}, \citealt{Babusiaux16review}, \citealt{Zasowski2016}) and the output of a numerical simulation which included gas/star formation and a major merger event \citep{Ath2016}. \citet{Debattista2019} used a cosmological simulation from the FIRE\footnote{https://fire.northwestern.edu/} project to study the vertex deviation as a function of age and metallicity in Baade's window. In agreement with the observations, they find that the high metallicity population has a large vertex deviation ($l_{v} \sim 40^{\circ}$) while it is negligible for metal poor stars. The variation of $l_{v}$ with age has not yet been studied in observations, but \citet{Debattista2019} find that the younger stars display a higher vertex deviation than older ones (their figure 10). The same $l_{v}$ trends with age and metallicity can be observed even if the accreted stars are not included, proving that they are not necessarly caused by an accreted population. 




In this work we aim to perform a quantitative comparison between observations and a self-consistent N-body simulation of bar formation, focusing on the links between bulge kinematics and bulge morphology. In particular, we study the relationship between the tilt of the velocity ellipsoid $l_{v}$ and the bar viewing angle $\alpha$. 
%

In Section \ref{section2}, we describe the data selection and the N-body boxy bulge model (\citealt{Shen2010}, thereafter the S10 model). In Section \ref{bulge}, we map the bulge velocity ellipsoids as seen in the data and the simulations while in Section \ref{fitting_method} we outline the fitting method and present the results. In Section \ref{metallicity} we add a new dimension to our 6D sample, the metallicity, and in Section \ref{conclusion} we present the conclusions.
%
%
\begin{figure*}
\hspace{-0.7cm}
\includegraphics[width=122mm]{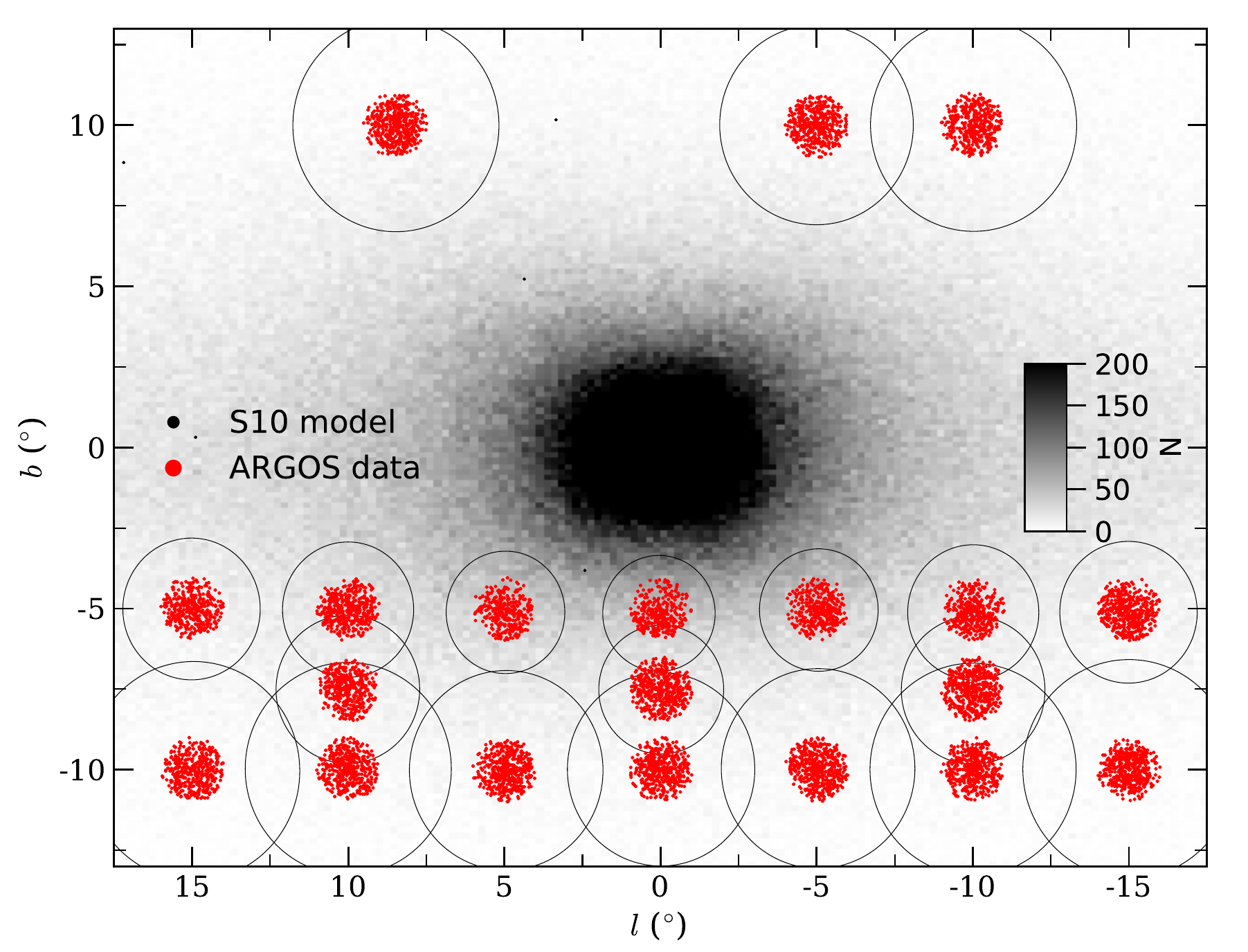}
\hspace{-0.4cm}
\includegraphics[width=63mm]{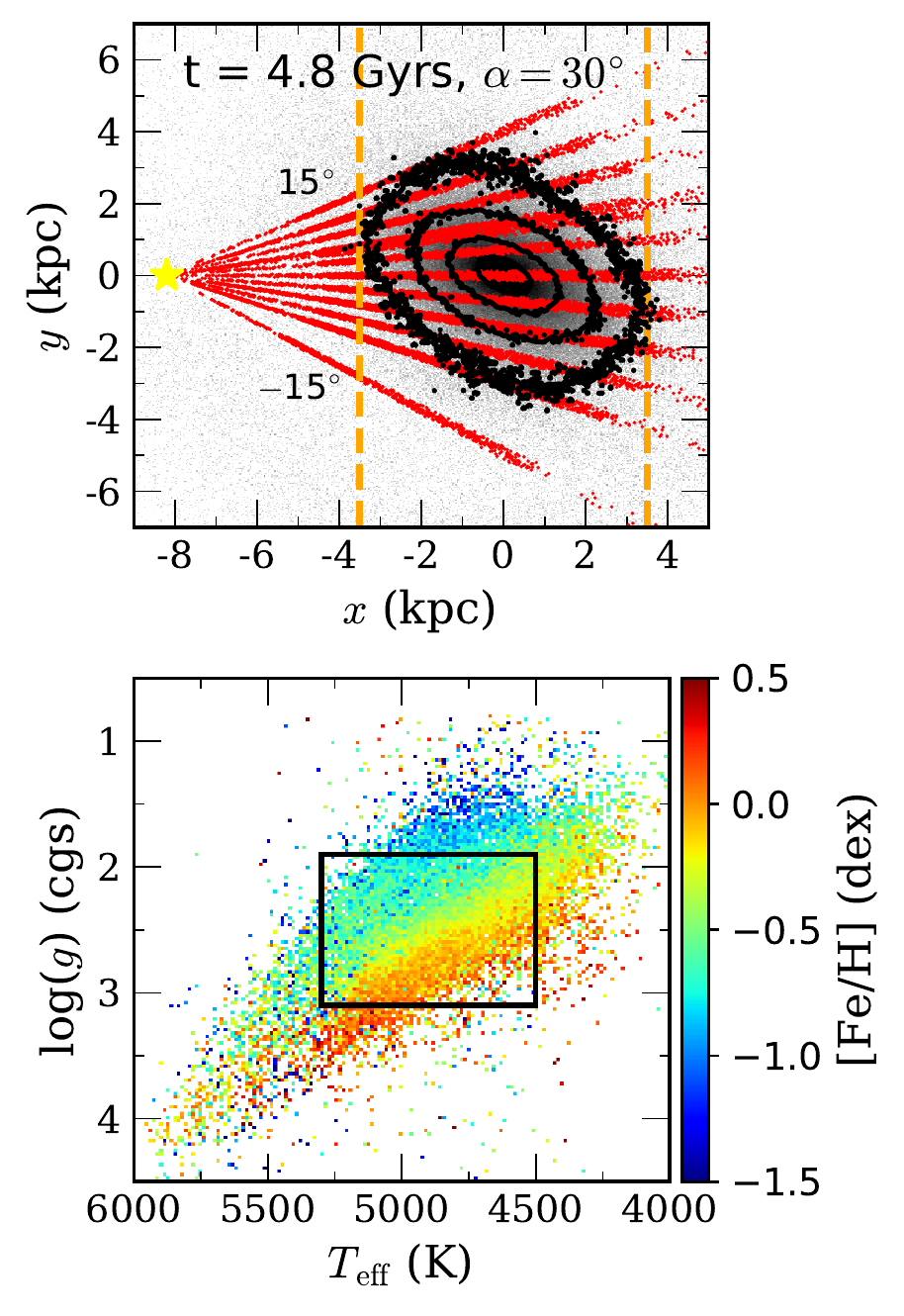}
\caption{\textit{Left panel}: The ARGOS survey (red) and the S10 model (gray), in Galactic coordinates. The number of particles in the model decreases dramatically with distance from the Galactic plane: to select at least 2000 simulation particles around each ARGOS pointing, we vary the radius of the simulation fields of view, each centred on a survey pointing. \textit{Top right panel}: Projection onto the $x$-$y$ plane of the ARGOS dataset and the simulation for a bar viewing angle of 30$^{\circ}$. In this configuration, the Sun is located at $x_{\odot}$ = -8.3 kpc, $y_{\odot}$ = 0 kpc and the positive longitudes are in the direction of positive $y$. The near end of the bar is at $x$ $<$ 0 kpc and $l > 0^{\circ}$. To minimize contamination from foreground stars, we only select stars within |$x$| $ < 3.5$ from the Galactic centre (vertical orange lines). \textit{Bottom right panel}: H-R diagram of the ARGOS stars with $Gaia$ DR2 proper motions. Stars within the black box are likely RC stars. To build a clean sample of bulge RC stars with 6-D phase-space information, we also perform proper motion error cuts in addition to the $x$ distance, log($g$) and $T_{\mathrm{eff}}$ selection. }
\label{lb}
\end{figure*}
\section{Data and the N-body Boxy Bulge Model}
\label{section2}
\subsection{Data}
The data originates from two surveys, ARGOS and $Gaia$ DR2.
\subsubsection{ARGOS}
ARGOS is a spectroscopic survey of 28000 predominantly giant stars in 28 fields \citep{Freeman2013, Ness2013kinematics}, selected for follow-up from the Two Micron All Sky Survey \citep[2MASS,][]{2mass}, in the magnitude range $K$ = 11.5 to 14 mag. The observations were taken with the AAOmega multi-fiber spectrograph on the Anglo Australian Spectrograph at the Siding Spring Observatory, which can observe up to 340 stars simulatenously. We are interested in studying the bulge kinematics therefore we focus only on the 20 fields (marked in red in Fig. \ref{lb}, left panel) that are closest to the main bulge population. The fields have a diameter of $\sim2^{\circ}$ and in each field around 1000 stars were randomly observed.

The radial velocity $v_{r}$, effective temperature $T_{\mathrm{eff}}$,  surface gravity log($g$), metallicity \textrm{[Fe/H]} and alpha element abundace [$\alpha$/Fe] were determined for each star using the ARGOS stellar pipelines \citep{Ness2012}. Radial velocities were computed via cross-correlation with synthetic spectra and, at the ARGOS typical resolution of $R=11,000$ and S/N $\sim$ 50 - 80, the velocity errors are smaller than 1.2 km/s \citep{Freeman2013}. In the following sections we assume a constant value of $\delta_{v_{r}} = 1$ km/s. 

 The distances were computed for the whole ARGOS sample via isochrone fitting \citep{Ness2013abundances}, but we choose to work only with a subsample of RC stars as they are great distance indicators and possess smaller distance uncertainties. The RC stars are selected based on their temperature and surface gravity, $4500$ <$T_{\mathrm{eff}}$/K$<5300$, 1.9 $<$log($g$)$<$ 3.1, as marked by the black lines in the ARGOS  Hertzsprung$-$Russell (H-R) diagram (bottom right panel of Fig. \ref{lb}; see also figures 2 and 3 from \citealt{Ness2013abundances}). Despite these cuts, the contamination from the background population of red giant branch (RGB) stars could be up to 30\% \citep{Freeman2013}.  It is difficult to separate the RC from the RGB but in this selection box, centred on the RC, they should have similar intrinsic brightness (see a model intrinsic luminosity curve $M_{\mathrm{K}}$ for the bulge giants in \citealt{Simion2017}, fig. 3). For the stars that are not on the RC, the $M_{\mathrm{K}}$ values were derived using isochrone fitting \citep{Freeman2013}. Reassuringly, we find a very close agreement between the distances provided by ARGOS and the distances computed directly from the extinction corrected photometry using the absolute magnitude value of the RC, $M_{K} \sim -1.61$ mag \citep{Alves2000, Hawkins2017} for our selected sample. The largest source of uncertainty is the spread of the RC absolute magnitude, $\delta_{M_{K}} \sim 0.22$ mag \citep{Alves2000, Ness2013abundances}, which gives uncertainties $\lesssim$1.5 kpc at the bulge distances. The errors due to 2MASS photometry and interstellar reddening are small at the ARGOS survey latitudes of $|b| > 4.5^{\circ}$. 

In the top right panel of Fig. \ref{lb}, we show the projection of the ARGOS RC stars onto the $x$-$y$ plane, where the Sun is placed at $(x_{\odot}, y_{\odot}, z_{\odot})$= (-8.3, 0, 0) kpc \citep{Gillessen2017}. While \citet{gravity} found $x_{\odot} = 8178 $ pc, we don't expect the small difference to impact our study, as the RC distances and transverse velocities dominate the uncertainties. We adopt a left-handed Galactic Cartesian system with the $x$-axis positive in the direction of the Galactic center, $y$-axis oriented along the Galactic rotation and the $z$-axis directed towards the north Galactic pole. In the following analysis, we only select stars within $|x| < 3.5$ kpc (orange lines in top right panel of Fig. \ref{lb}) from the Galactic Centre (GC) $(x, y)$ = (0, 0) kpc in order to minimize contamination from disc and foreground stars. 
\begin{figure}
\hspace{-0.5cm}
\includegraphics[width=90mm]{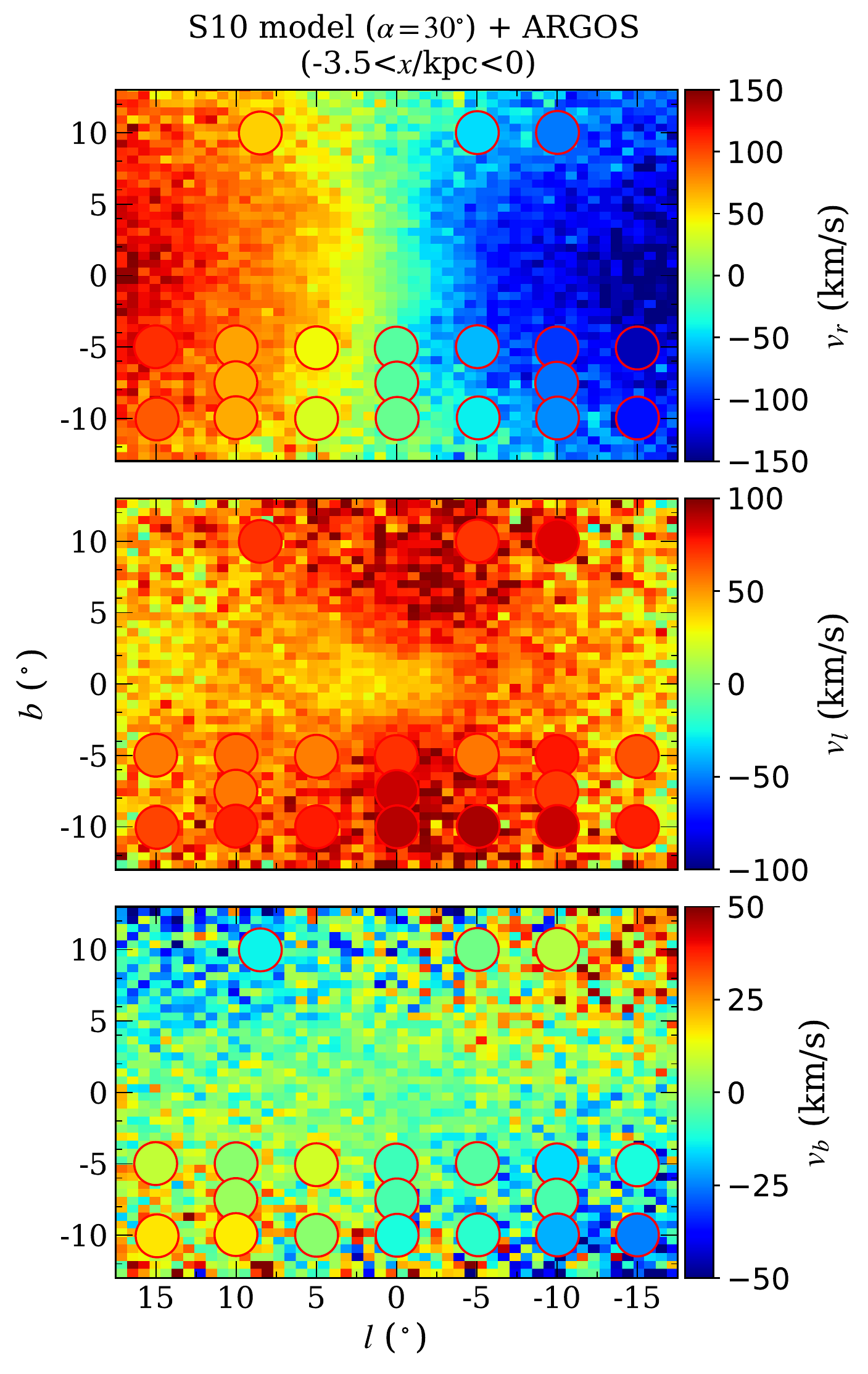}
\caption{\textit{Top panel}: Radial velocity map of the S10 model and, overlaid, the median radial velocity in each ARGOS pointing, in Galactic Coordinates. Only stars with $-3.5<x$/kpc$<0$ (in front of the GC) are shown, to facilitate the comparison between the simulation and the data, which are less complete behind the GC. \textit{Middle panel}: same as above, but for $v_{l}$. Stars in front of the GC, due to the bar's rotation, move from left (negative longitudes) to right (positive longitudes) causing $v_{l}$ to be positive. In addition, there is little variation with latitude because the bulge rotates approximatively cylindrically. \textit{Bottom panel}: Same as above, but for $v_{b}$. The vertical motion is small (notice the colour-scale change) with $|v_{b}| \lesssim 50$ km/s.}
\label{vels}
\end{figure} 
\subsubsection{GAIA DR2}
$Gaia$ DR2 provides accurate proper motions measurements for the majority of ARGOS stars: from the initial ARGOS sample, we discard targets which do not have a $Gaia$ DR2 counterpart or have large proper motions uncertaintities $\sigma_{\mu_{\mathrm{RA}}}, \sigma_{\mu_{\mathrm{Dec}}} >$ 0.2 mas/yr. The cross matching between ARGOS and the 2MASS - $Gaia$ DR2 value added catalog was done within a $1\arcsec$ radius but, after applying the proper motions error selection, all matches were within $0.3\arcsec$ with a mean angular distance of 0.05\arcsec. We have also checked that the K magnitudes in the ARGOS and 2MASS - $Gaia$ DR2 catalogs were matching.
\begin{figure*}
\hspace{-0.85cm}
\includegraphics[width=194mm]{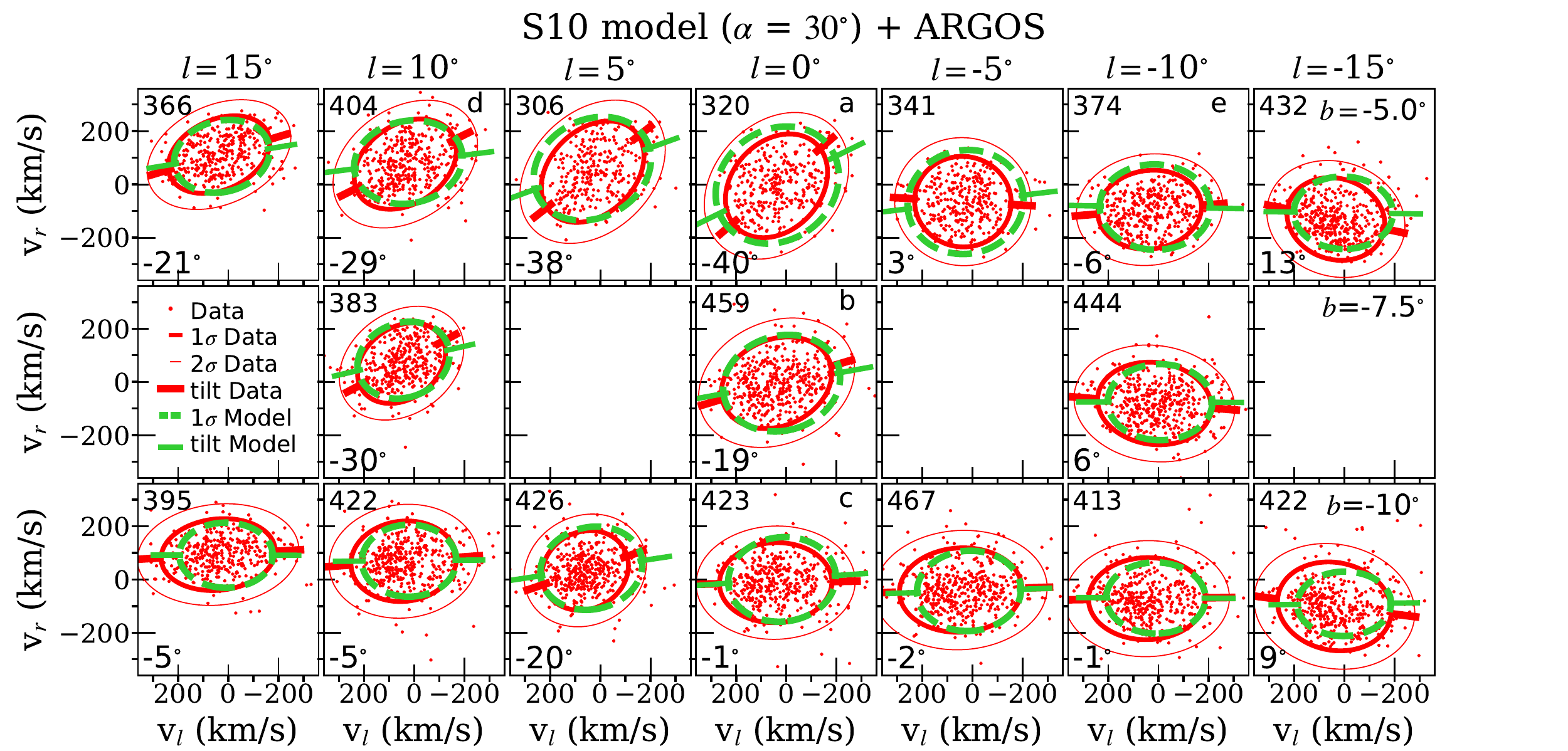}
\caption{Velocity ellipsoids for the ARGOS survey (each panel corresponds to a pointing in Fig. \ref{lb}). The 68\% and 95\% contours of the ARGOS velocity distributions are shown in red, and the  68\% contour of the distance-resampled S10 model (see Fig. 
\ref{test}) with a viewing angle of 30$^{\circ}$, in green. The tilt of the ellipsoids $l_{v}$ is indicated by the red/green lines which extend between the 68\% and 95\% contours of the data/simulation velocity ellipsoids. The number of ARGOS RC bulge stars (red points) in each panel is given in the top-left corner and the corresponding $l_{v}$ value in the bottom-left corner. The $l_{v}$ values of the distance-resampled S10 model and the data are also shown in Fig. \ref{vertex_south}.}
\label{velsgrid}
\end{figure*} 
Finally, our sample of bulge RC stars with complete 6D phase space information amounts to $\sim$ 7,000 stars, or around 400 stars per pointing.

In Fig. \ref{vels} we show the median of the 3 velocity components in each ARGOS field from Fig. \ref{lb}: the line-of-sight velocity $v_{r}$ (top panel), the longitudinal velocity $v_{l}$ (middle panel) and latitudinal velocity $v_{b}$ (bottom panel). Only stars in front of the GC ($-3.5<x/$kpc $<0$) are shown because the ARGOS sample is more complete at nearby heliocentric distances than behind the GC. The velocities were corrected for the Solar reflex motion assuming the default $\texttt{astropy}$ values for the Sun's peculiar motion, (U, V, W) = (11.1, 12.24, 7.25) km/s \citep{Sch10} and for the circular velocity at solar radius, 220 km/s. From these maps, it is immediately clear that the stars do not have random motions: the stars at positive/negative longitudes have positive/negative radial velocities respectively resulting from a perpective effect of the approaching right-side versus retreating left side of the bar (see also \citealt{Ness2013kinematics}). The $v_{l}$ velocities are all positive, as expected for stars in front of the bar (\citealt{Qin2015}, fig. 4). The $v_{b}$ velocities are small across the bulge, within $|v_{b}| <$ 50 km/s, compared to the values of $v_{r}$ and $v_{l}$.

Beyond |l| $>$ 10$^{\circ}$, the central boxy-peanut bulge of the MW transitions to a longer, flatter bar which extends out to $l\sim25^{\circ}$ \citep[e.g.][, fig. 9]{Wegg2015}; however, being limited by the survey to fields beyond $|b|>4.5^{\circ}$, the long (thin) bar is not visible in our $l \sim \pm10^{\circ}$, $\pm15^{\circ}$ fields. At $l\sim20^{\circ}$, the outermost longitude of ARGOS which is not considered in this work, the long bar lies at a distance of $\approx$ 5.2 kpc from the Sun and a height above the plane of 180 pc, still well below the ARGOS visibility threshold of $\sim$400 pc at this distance. Therefore, we can safely assume that the long bar does not affect the kinematics observed in the fields considered in this work, and we do not discuss it in the next sections.
\subsection{Simulations}
We use the S10 model, an N-body simulation with 1 million disc particles rotating in a rigid dark matter potential. In this model, a bar is formed in the early stages of evolution which buckles to produce a boxy peanut shaped bulge. The S10 model is successful at reproducing and explaining some of the observed morphological bar properties such as the double red clump, X-shape and kinematics \citep{Li2012, Molloy2015clumping, Molloy2015orbits, Nataf2015, Qin2015}. The simulation does not include gas and star formation therefore is expected to characterise the kinematics of the older bulge stellar population. As newly formed stars reside closer to the Galactic Plane \citep{Debattista2015} and ARGOS is limited to |b|$>4.5^{\circ}$, we are not concerned with the highly star forming disc regions. 

Strong variations in the bar pattern speed $\Omega_{\mathrm{P}}$ (of the order of $\pm10$  km/s/kpc) are expected to have an effect on the bulge mean radial velocities and velocity dispersions (see \citealt{Portail2017}, fig. 10). However, the bar pattern speed in the S10 model, $\Omega_{\mathrm{P}} \sim 40 $ km/s/kpc \citep[][]{Shen2014}, matches closely the MW value measured from kinematics: \citet{Portail2017} computed $\Omega_{\mathrm{P}} = 39 \pm 3.5$ km/s/kpc using ARGOS data and, more recently, \citet{Sanders19} found $\Omega_{\mathrm{P}} = 41 \pm 3 $ km/s/kpc using proper motions from $Gaia$ DR2 and VVV. Because the bar pattern speed $\Omega_{\mathrm{P}}$ of the model closely matches the observations, we do not expect it to be the cause of significant kinematic differences between the data and the model.

A first qualitative comparison between our data and the S10 model kinematics is shown in Fig. \ref{vels}, within the $-3.5<x$/kpc$<0$ distance range (in front of the GC). For the S10 model we assume a bar viewing angle of $\alpha =30^{\circ}$. The three panels of the figure are maps of the median velocity values of the three velocity components in the S10 model. The median velocities in each ARGOS field are shown with colour-coded circles (see Sec. 2.1). The S10 maps look similar for different bar viewing angles $\alpha$ within this distance range. If we considered the region with $0<x$/kpc$<3.5$ (behind the GC), only the $v_{l}$ map would change significantly, as stars at these distances have $v_{l} < 0$ km/s. Overall, the velocity trends observed in the data and the S10 model are consistent. In the next sections, we perform a quantitative comparison between the two. 
\section{Bulge velocity ellipsoids}
\label{bulge}
\subsection{Data}
The ARGOS velocity distributions in the radial-longitudinal velocity plane are shown in Fig. \ref{velsgrid}, where each subpanel corresponds to a survey pointing, shown in red in Fig. \ref{lb}, where we exclude the three fields with $b > 0^{\circ}$. The near-end of the bar is situated at positive longitudes, in the left hand-side of the figure, while the far-end is situated at negative longitudes in the right hand-side. For any line of sight, we model the velocity distributions along the longitudinal ($l$), radial ($r$) and latitudinal ($b$) velocity directions with a single-component 3-dimensional Gaussian. We assume that the distributions in each pointing are Gaussian although we might expect deviations from Gaussianity due to incomplete sampling and large measurement uncertaintities for the stars on the far side of the bar. To compute the model likelihood we use the Extreme Deconvolution \citep[ED,][]{ED} method implemented in the $\mathrm{astroML}$ \citep{astroML} package. In Fig. \ref{velsgrid} we show the contours containing 68\% (thick red line) and 95\% (thin red line) of the ARGOS velocity distributions fitted in each individual field. The number of RC stars that satisfy all the selection criteria outlined in Section \ref{section2} and that are used to fit the velocity ellipsoid, is given in the upper left corner of each subpanel. 

The ED method requires that the uncertainty of each velocity component  $\delta_{V_{i}} = \{\delta_{v_{l}}, \delta_{v_{b}}, \delta_{v_{r}} \}_{i}$ is provided. For each star we have the 6D phase-space information (see Sec. 2) provided by the $Gaia$ - ARGOS cross-match $\{\alpha, \delta, D, \mu_{\alpha^{*}}, \mu_{\delta}, v_{r} \}_{i}$. The uncertainties $\delta_{V_{i}}$ are computed via Monte-Carlo re-sampling where the diagonal terms of the covariance matrix are the $Gaia$ DR2 errors on the right ascension $\sigma_{\alpha^{*}}$, declination $\sigma_{\delta}$ and proper motions  $\sigma_{\mu \alpha^{*}}$, $\sigma_{\mu \delta}$ and the ARGOS heliocentric distance uncertainties $\sigma_{D}$ and radial velocity error of 1 km/s. The $Gaia$ cross-terms between the coordinates $\rho(\alpha, \delta)$ and proper motion components $\rho(\mu_{\alpha^{*}},  \mu_{\delta})$, including $\rho(\alpha,  \mu_{\delta})$, $\rho(\delta, \mu_{\alpha^{*}})$, are also taken into account. We use the standard deviation of 1000 evaluations of $V_{i} = \{v_{l}, v_{b}, v_{r}\}$ as an estimate of the star's velocity uncertainty $\delta_{V_{i}}$. Because the uncertainties on the $v_{l}$ and $v_{b}$ components are dominated by the distance errors, our most uncertain measurement, they can reach $\sigma_{v_{l}},  \sigma_{v_{b}} \sim 30-40$ km/s. 
\begin{figure}
\hspace{-0.5cm}
\includegraphics[width=90mm]{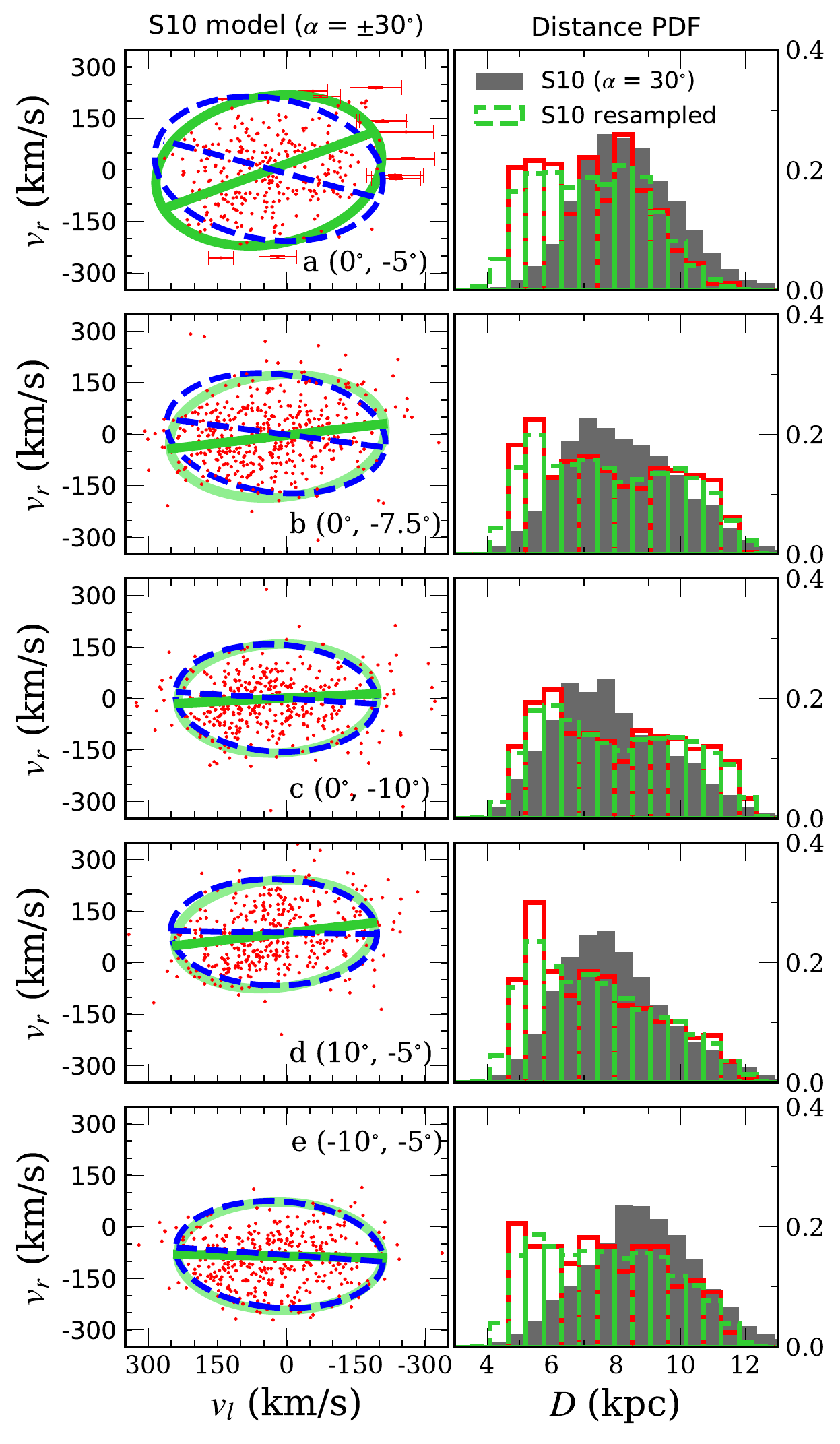} 
\caption{\textit{Left panels:} To find the most likely bar viewing angle from the ARGOS velocity distribution, we compute the probability of the data (red points) to belong to a model velocity ellipsoid fitted on the distance-resampled S10 model. We show the model for 2 bar angles, $\alpha = $-30$^{\circ}$ (blue) and $\alpha = $30$^\circ$ (same as in Fig. \ref{velsgrid}, green). For one field, we have added the uncertainties on the $v_{l}$ component for stars outside the 68\% contour. The errors vary between $15$ to $50$ km/s, with larger values for the stars behind the bar, which have $v_{l} < 0$ km/s. In the fields labeled $c$, $d$ and $e$, the vertex deviation is $l_{v} \approx 0^{\circ}$ for both models, indicating that these fields are not good predictors of the bar viewing angle. On the other hand, the model velocity ellipsoids and the tilt of their major axis, are different in the a, b fields, for the two angles. \textit{Right panels:} The distance distribution of the simulation particles is centred on the GC (black histogram) but the data tends to have more stars in front of the bar (red histogram) due to the survey sampling. The models shown in the left column correspond to a resampled distance probability distribution (green histogram), which matches the data distance distribution. }
\label{test}
\end{figure} 
\begin{figure}
\includegraphics[width=80mm]{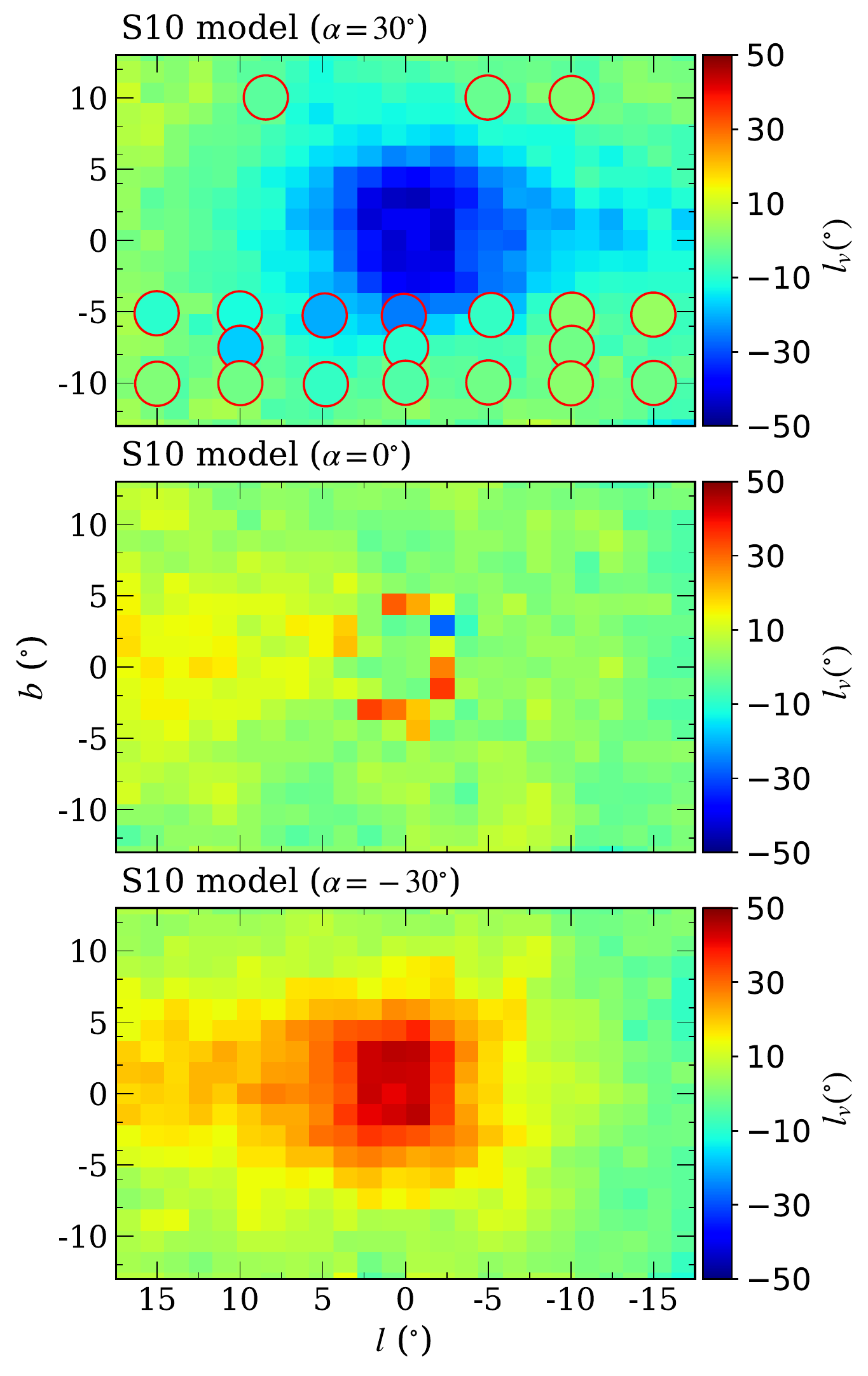}
\caption{The $l_{v}$ map of the S10 model for 3 viewing angles: $30^{\circ}$ (\textit{top panel}), where the near end of the bar is located at positive longitudes as in the MW; $0^{\circ} $ (\textit{middle panel}) where the bar major-axis is alligned with the Sun-GC line of sight; $-30^{\circ}$ (\textit{bottom panel}), where the near end of the bar is located at negative longitudes. All the particles in the simulation with $|x|< 3.5$ kpc were used for these maps. The $l_{v}$ map changes when the distance distribution of the simulation is resampled to match the ARGOS distances distribution, as shown in the top panel. The resampled particles were selected from circular fields (red circles in Fig. \ref{lb}) centred on the ARGOS fields.}
\label{vertex}
\end{figure} 
\subsection{Simulations}
 For each particle in the S10 model, the 6D phase-space $\{x,y,z, v_{x}, v_{y}, v_{z}\}_{i}$ is provided in the Galactocentric frame. The transformations to the Galactic frame were performed using $\texttt{galpy}$. 
 The advantage of using a simulation is that we can adopt any angle between the Sun-GC line and the bar major axis, $\alpha$, by rotating the reference frame. The simulation particles are selected from circular fields of varying radius, centred on the ARGOS fields. As the density of particles drastically diminishes with distance from the Galactic Centre, we increase the radius of the simulation fields with increasing longitude and latitude (black circles in Fig. \ref{lb}), so that each field contains approximately 2000 particles within $|x|<3.5$ kpc. From Fig. \ref{lb} it is clear that the simulation fields at  $b = \pm 10^{\circ}$ are significantly bigger than the corresponding ARGOS fields but we do not expect a small difference in field size to have a major effect on the kinematics of stars at these latitudes as they are situated on the outskirts of the main bulge population (see the number density map in gray, in the left panel of Fig. \ref{lb}).
 
In Fig. \ref{velsgrid}, we show the 68\% contour (green dotted line) of the S10 model velocity distributions within each simulation field, for $\alpha = 30^{\circ}$. The model closely matches the data 68\% contour (red line) in most fields. The distance distribution of the simulation particles was resampled according to the distribution of the ARGOS distances, with a process detailed in Fig. \ref{test}. In the left column of the figure we show the model for two $\alpha$ values and in right column we show the distance distribution of the simulation particles before resampling (black histogram) in 5 fields: 3 along the minor-axis (labelled a, b, c), one on the near end of the bar (d) and one on the far side of the bar (e). While the S10 model particles are concentrated around the GC as expected from star count models \citep[e.g.][]{Simion2017}, the distribution of the ARGOS distances (red histogram) is skewed, with the majority of stars located in front of the GC, likely due to a mismatch between the S10 model density and the MW bulge density distribution, incomplete survey sampling, extinction and magnitude limits. The distance to the stars is important as the stellar kinematics varies within the bar: for example, the 2D velocity distribution of ARGOS stars (red points) in Figures \ref{velsgrid} and \ref{test} shows there is a higher concentration of stars for $v_{l} > 0$ km/s values than for $v_{l} < 0 $ km/s, which is to be expected if the majority of stars is in front of the bar \citep{Qin2015}. Therefore, to build a model that best describes the data, we resample the particles in the simulation to follow the same distance distribution as the ARGOS RC stars in each field.

Before resampling, to mimick the observational procedure, we randomly perturb the heliocentric distances $D$ in the simulation, which are unaffected by errors, by the typical uncertainties expected for RC stars of $\delta_{M_{K}} \sim 0.22$ mag assuming a Gaussian error distribution with a standard deviation of $\sigma_{D} \approx \delta_{M_{K}} \times 0.2 \times D$ $\times$ ln(10). From the perturbed sample we draw 25000 random particles, allowing for duplicates, according to the probability distribution of the ARGOS distances (red histogram in Fig. \ref{test}), modelled with a kernel density estimation (KDE) in each individual pointing. The probabily density distribution of the resampled S10 model distances is shown with a green histogram in Fig. \ref{test} and it closely matches the data, red histogram, by construction. 
 \begin{figure}
\hspace{-0.5cm}
\includegraphics[width=90mm]{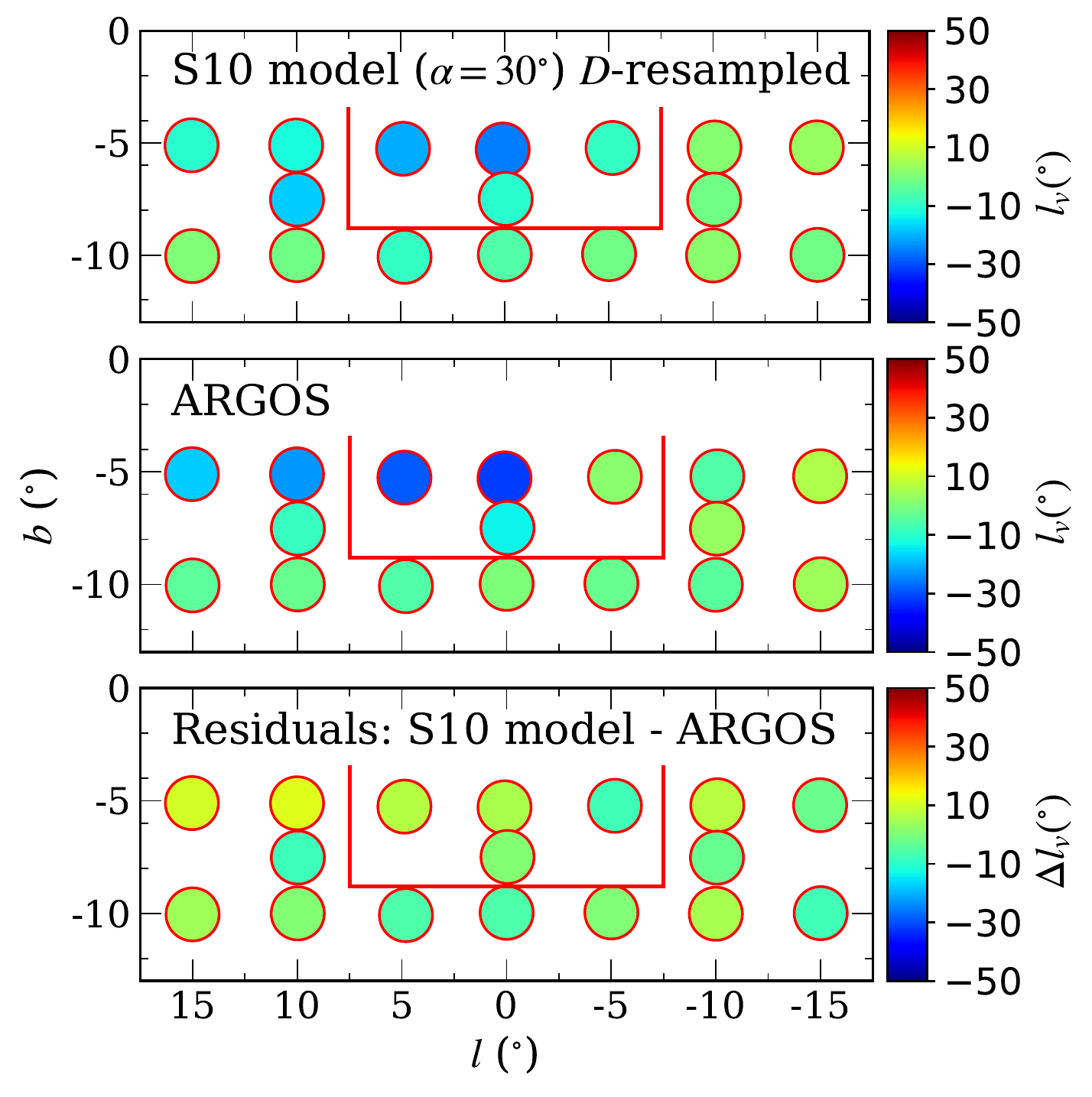}
\caption{\textit{Top panel:} Same as in top panel of Fig. \ref{vertex}. The $l_{v}$ values were computed using only the particles within the red circles in Fig. \ref{lb}. The simulation was resampled so that the heliocentric distance $D$ distribution matches the ARGOS data sampling. \textit{Middle panel:} The data $l_{v}$ map is in good agreement with the resampled simulation, in particular in the 4 central fields delimited by the red lines. \textit{Bottom panel:}  The difference between the resampled simulation and data $l_{v}$ shows no systematic trends.}
\label{vertex_south}
\end{figure} 

We also add realistic velocity errors to the simulation particles. In each field of view, we assign the median ARGOS velocity uncertainty of stars in that field, to each S10 model particle. The 3D velocity distribution of these particles is then fitted with a multivariate Gaussian, for a specified viewing angle $\alpha$, in each individual pointing. In the right column of Fig. \ref{test} we show the 68\% contours of two models, one with the near end of the bar at positive longitudes $l>0^{\circ}$ and $\alpha = 30^{\circ}$ (green) and one with the near end of the bar at negative longitudes $l<0^{\circ}$ and $\alpha = -30^{\circ}$ (dotted blue line). The model in Fig. \ref{velsgrid} (green line, $\alpha =  30^{\circ}$) is identical to the one in Fig. \ref{test} and is also computed using the distance-resampled S10 model. 
\subsection{Vertex deviation map}
\label{devmap}
Another quantity that is useful in describing the velocity ellipsoid is the vertex deviation $l_{v}$  (\citealt{Zhao1994}), the angle which measures the tilt of the longest axis of the velocity ellipsoid in the $v_{r}$ - $v_{l}$ plane:
\begin{equation}
l_{v} = \frac{1}{2}\mathrm{arctan} \bigg(  \frac{2 \sigma^{2}_{rl}}{|\sigma^{2}_{rr} - \sigma^{2}_{ll} |} \bigg),
\label{vertexequation}
\end{equation} 
where $\sigma_{rl}$, $\sigma_{rr}$ and $\sigma_{ll}$ are the convariance and standard deviation terms of the velocity components along the line-of-sight ($r$) and longitude ($l$) directions. By definition, $l_{v}$ takes values between -45$^{\circ}$ and +45$^{\circ}$.

We map the $l_{v}$ variation in the simulation for three bar angles  $\alpha = 30^{\circ}$ in Fig. \ref{vertex} (top panel), $\alpha = 0^{\circ}$ (middle) and $\alpha = -30^{\circ}$ (bottom) using all particles in the S10 model within $|x|<$3.5 kpc. The maps provide clear evidence that $l_{v}$ is strongly dependent on the bar viewing angle $\alpha$, especially in the fields close to the GC, $|b|\lesssim 7.5^{\circ}$. In addition, for a given $\alpha$, $l_{v}$ is not constant across the bulge as already suggested by the velocity trends in Fig. \ref{vels}. 

The vertex deviation values computed with the ARGOS-$Gaia$ DR2 sample are in disagreement with the S10 model predictions for $\alpha = 30^{\circ}$ before resampling (top panel of Fig. \ref{vertex}) but in good qualitative agreement after distance-resampling (Fig. \ref{vertex_south}). The residuals between the distance-resampled S10 model  $l_{v}$ values and the ARGOS data (bottom panel of Fig. \ref{vertex_south}) are close to $0^{\circ}$ in most fields and show no systematic trends, proving that resampling is a fundamental step in building the S10 model velocity ellipsoid.

\begin{figure*}
\includegraphics[width=180mm]{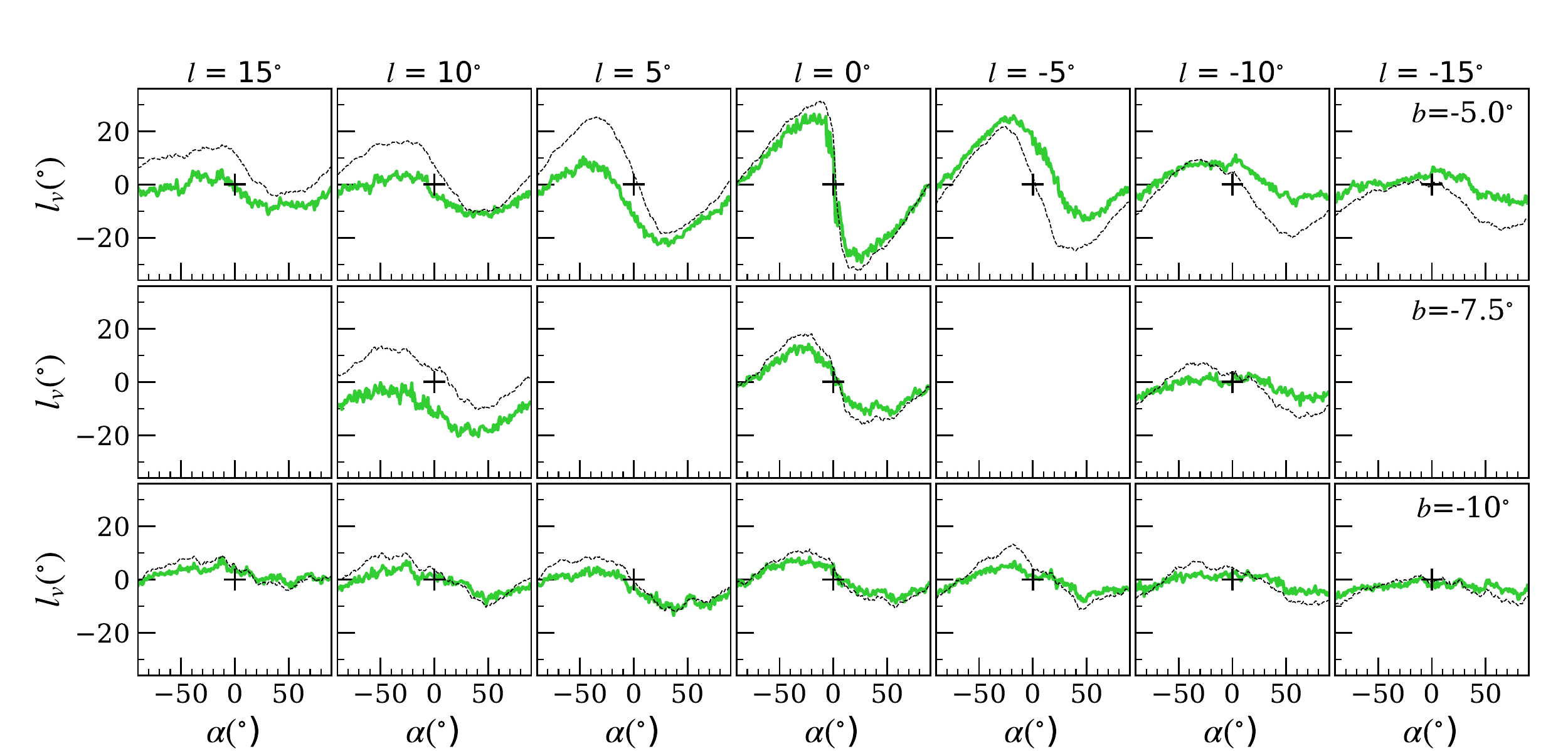}
\caption{Vertex deviation of the resampled (green line) and non-resampled (black dotted line) model as a function of bar viewing angle, $\alpha$, for 180 values between $-90^{\circ}$ and $+90^{\circ}$. For $\alpha = 0^{\circ}$, the vertex deviation is $l_{v} = 0^{\circ}$ (see black cross), therefore an axisymmetric density distribution, would not generate a tilted velocity ellipsoid, as it can be seen also in the middle panel of Fig. \ref{vertex}. In the fields close to the GC the vertex deviation variation at $\alpha = 0^{\circ}$ is very steep, which causes the circular artifact at $|l|, |b| < 5^{\circ}$. }
\label{vertex_alpha}
\end{figure*}

In Fig. \ref{vertex} we mapped the S10 model vertex deviation for $\alpha = 0^{\circ}$, $\pm 30^{\circ}$ and showed that the bar viewing angle has a direct influence on kinematics (hence on $l_{v}$). In Fig. \ref{vertex_alpha}, we show the variation of $l_{v}$ with $\alpha$ for a grid of values between $-90^{\circ}$ and $+90^{\circ}$, for both the resampled (green curve in Fig. \ref{vertex_alpha}) and the non-resampled (black curve) simulation. The black curve passes through $(\alpha$, $l_{v}) = (0^{\circ}, 0^{\circ})$ marked with a black cross in the figure, confirming that for an axisymmetric system the vertex deviation is null across the bulge. In addition, the amplitude of the $l_{v}$ variation is stronger for the non resampled model (black curve) than for the resampled model (green curve), suggesting that the ARGOS distance sampling will slightly affect the $l_{v}$ measurements. In particular, the resampling affects $l_{v}$ in the low latitude fields at $b=-5^{\circ}$ except for the most central field $(l, b) =$ $(0^{\circ}, -5^{\circ})$ which displays the strongest $l_{v}$ variation with $\alpha$ with values between +$35^{\circ}$ and $-35^{\circ}$ for both the distance-resampled and non-resampled model. Slightly smaller $l_{v}$ variations of $\pm25^{\circ}$ can be seen in the three adjacent fields $(l, b) = \{(5^{\circ}, -5^{\circ}), (-5^{\circ}, -5^{\circ}), (0^{\circ}, -7.5^{\circ})\}$ (delimited by a red box in Fig. \ref{vertex_south}) but they sharply decrease beyond that (Fig. \ref{vertex_alpha}). This is confirmed by Fig. \ref{test} where the S10 model velocity ellipsoids in three fields  c (0$^{\circ}$,-10$^{\circ}$), d (-10$^{\circ}$,-5$^{\circ}$) and e (10$^{\circ}$, -5$^{\circ}$)  almost overlap for $\alpha=30^{\circ}$ and $\alpha=-30^{\circ}$ and $l_{v}$ is almost identical, a strong indication that beyond the 4 central fields, the kinematics (and $l_{v}$) is not sensitive to the bar viewing angle. 

Our measurements along the minor axis are consistent with the results from \citealt{Soto2012} who obtained $l_{v}$ $\sim$ -43$^\circ$/40$^{\circ}$ at  (1$^\circ$, -4$^\circ$) for all stars/RGBs and -17$^\circ$ at (0$^\circ$, -6$^\circ$) for all stars. We have obtained remarkably similar results, $l_{v} = -40^{\circ}$ at (0$^\circ$, -5$^\circ$) and $l_{v} = -19^{\circ}$ at (0$^\circ$, -7.5$^\circ$). Both studies agree that the $l_{v}$ values decrease at increasing latitudes and longitudes, away from the Galactic Centre.  

In the next section, we provide a more quantitative comparison between the data and the simulation based on the velocity vectors of the individual stars.   
\section{Retrieving $\alpha$ from kinematics}
\label{fitting_method}
We aim to constrain the angle between the MW bar major-axis and the Sun-GC line $\alpha$, a fundamental property of the MW bar morphology from stellar kinematics. 
\subsection{Kinematic modelling}
We determine the most probable bar angle from kinematic observations via a maximum likelihood method. The probability density in velocity space of a bulge RC star defined by its three velocity components $V_{i} = \{v_{l}, v_{b}, v_{r}\}_{i}$ to belong to a model $M(\mu, \Sigma'; \alpha)$, fitted on the S10 model velocity distribution after distance-resampling, is given by the Gaussian probability density function:
\begin{equation}
p(V_{i}|\alpha) =  \frac{1}{( 2 \pi )^{n/2} | \Sigma' | ^{1/2}} \mathrm{exp} \bigg(\frac{1}{2} (V_{i} - \mu)^{T} \Sigma'^{-1} (V_{i} - \mu)  \bigg)
\end{equation}
where $n = 3$ is the size of the data vector $V_{i}$, $\mu$ is the centroid of the velocity ellipsoid, and $\Sigma'$ the total covariance matrix $\Sigma' = \Sigma + \Sigma_{\mathrm{D}}$ where $ \Sigma$ is the 3 x 3 covariance matrix of the velocity ellipsoid
\[ \Sigma = \left[ \begin{array}{ccc}
\sigma_{ll}^2 & \sigma_{lr}^2 &  \sigma_{lb}^2   \\
 \sigma_{rl}^2  & \sigma_{rr}^2 &  \sigma_{rb}^2  \\
 \sigma_{bl}^2  &  \sigma_{br}^2  & \sigma_{bb}^2\end{array} \right].\] 
The diagonal terms are the velocity dispersions along the three directions $\sigma_{ll}, \sigma_{rr}, \sigma_{bb}$  and the cross terms  $\sigma^{2}_{lr}, \sigma^{2}_{br}, \sigma^{2}_{lb}$ determine the orientation of the velocity ellipsoid. $ \Sigma_{\mathrm{D}}$ is a diagonal matrix which contains the data uncertainties along the three velocity components, $\sigma^{2}_{v_{l}}, \sigma^{2}_{v_{r}}, \sigma^{2}_{v_{b}}$, computed using Monte Carlo resampling, as explained in Section 3.1. Both the centroid $\mu$ and the covariance matrix $\Sigma$ are computed on the S10 model and are $\alpha$ dependent. Throughout this work, we have abbreviated $\mu(\alpha)$ and $\Sigma(\alpha)$ with $\mu$ and $\Sigma$.

For each pointing,  the probability density of the observed sample of velocities $V$ under the model $M({\mu, \Sigma; \alpha})$ for the bar viewing angle $\alpha$, is given by:
\begin{equation*}
P(V| \alpha) = \prod\limits_{i=1}^N p(V_{i}| \alpha).
\end{equation*}
In practice, we aim to find $\alpha$, for which the quantity:
\begin{equation}
- \mathrm{ln} (P) = - \sum\limits_{i=1}^N \mathrm{ln} p(V_{i}| \alpha)
\label{lnL}
\end{equation}
is a minimum, where the sum is carried out for the total number of stars $N$, in each pointing.  The viewing angle $\alpha_{\mathrm{min}}$, which minimizes Eq. \ref{lnL}, is then the Maximum Likelihood estimate (ML).  The $1\sigma$ error on $\alpha_{\mathrm{min}}$ is defined by the interval $\Delta \mathrm{ln}(L) = 0.5$, above the minimum of the log-likelihood curve\footnote{see \cite{Bev03}}. 
%
\begin{figure*}
\includegraphics[width=180mm]{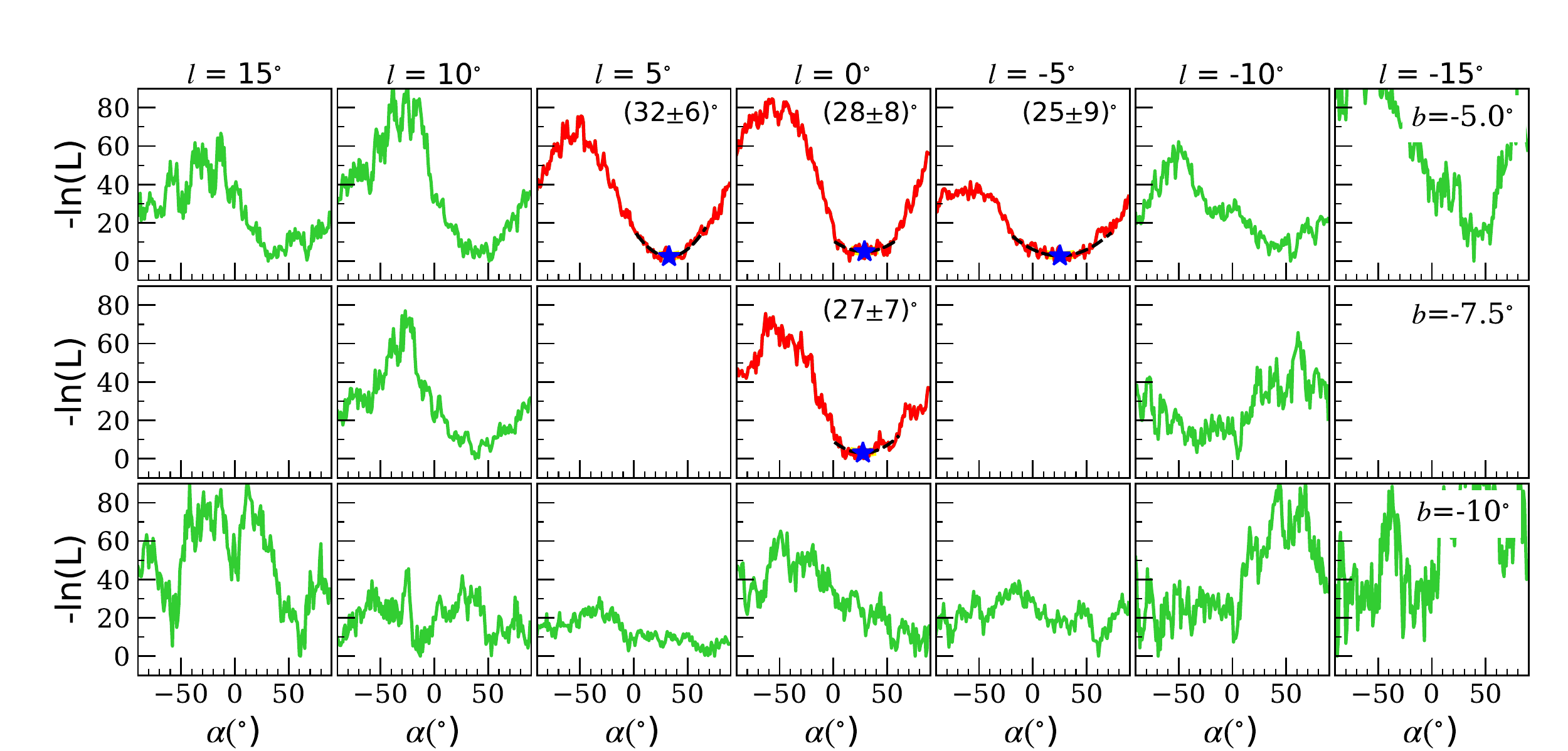}
\caption{The log-likelihood of the data computed using the model fitted on the resampled S10 model. The blue point is the best-fit angle in each field of view. We provide the $\alpha$ values for the 4 fields marked in red, and the uncertainties. }
\label{logL}
\end{figure*} 
\subsection{Results}
We have applied the fitting method to the individual fields and we show the log-likelihood variation with $\alpha$ in Fig. \ref{logL}. The log-likelihood was evaluated for 180 $\alpha$ values using the distance-resampled S10 model (the vertex deviation of this model for each angle is given in Fig. \ref{vertex_alpha} and is discussed in Section 3.3) and it reaches a minimum for different values of $\alpha$ in each field.  In the figure we placed the minimum log-likelihood at 0 in each pointing.
\begin{figure}
\hspace{-0.9cm}
\includegraphics[width=47.mm]{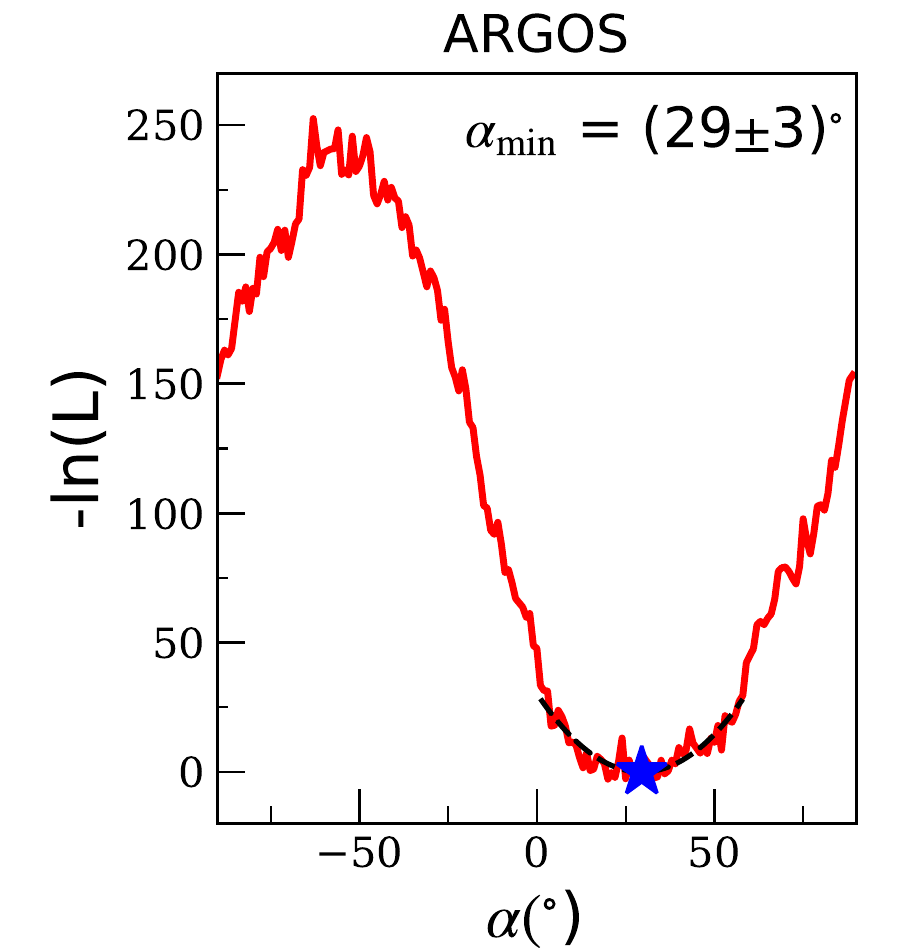} 
\hspace{-0.4cm}
\includegraphics[width=47.mm]{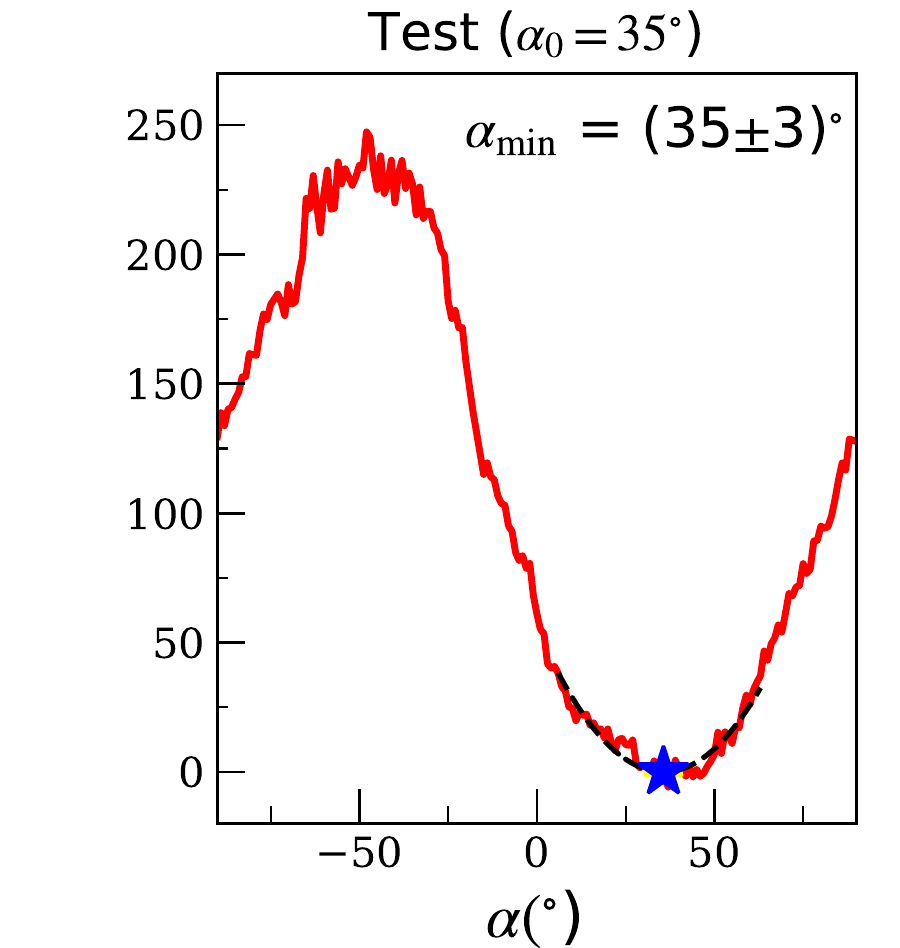} 
\caption{\textit{Left panel:} Final result obtained from the sum of the log-likelihood curves in the 4 fields marked in red in Fig. \ref{logL}. The minimum is obtained at $\alpha_{\mathrm{min}} = (29 \pm 3)^{\circ} $, where the uncertainty was computed over a 9$\sigma$ interval. \textit{Right panel:} We perfomed a test on a mock sample with $\alpha_{0}=35^{\circ}$ and the same distance distribution as the data. We have done the test using different angles for the mock sample, and each time we successfully recover the true angle, $\alpha_{0}$. }
\label{logLfinal}
\end{figure} 

 ARGOS is a survey at intermediate latitudes and it can only sample the outer edges of the bulge stellar density distribution. The fields $(l, b) = \{(5^{\circ}, -5^{\circ}), (-5^{\circ}, -5^{\circ}), (0^{\circ}, -7.5^{\circ})\}$ are the closest to the bulge center and contain a large proportion of bulge stars. Here, the stellar kinematics is most affected by the bar morphology as we have shown in Fig. \ref{vertex_alpha}: the 4 central fields exhibit large $l_{v}$ variations with $\alpha$  and have clear minimae while the outer fields have much smaller variations especially after resampling (green curve). For this reason (see also Sec 3.3), to determine the best-fit viewing angle $\alpha_{\mathrm{min}}$ we sum the log-likelihood curves only in the 4 central fields marked in red in Fig. \ref{logL}. The result is shown in the left panel of Fig. \ref{logLfinal}.

To determine $\alpha_{\mathrm{min}}$, a parabola was fitted over a $\Delta \mathrm{ln}(L) = 40.5$ (9$\sigma$) interval (black dashed line around the minimum). The final value is $\alpha_{\mathrm{min}}  = (29 \pm 3)^{\circ}$. This value is an independent measure of the MW bar viewing angle from the kinematics of $\sim$ 7000 ARGOS-$Gaia$ DR2 RC stars, and is consistent with previous studies of bar morphorlogy using star counts (\citealt{Stanek1997}, \citealt{Wegg2013}, \citealt{Cao2013}, \citealt{Simion2017} - see their fig. 17 for a comparison between different works).
\subsection{Validation tests}
\subsubsection{Mock sample}
We test whether we are able to recover the correct bar angle of a mock sample of particles drawn from the S10 simulation with the bar major-axis at a given angle $\alpha_{0}$ with the Sun-GC line, replicating the same fitting procedure applied to the data. 

The mock sample distances include realistic heliocentric distance errors, modelled assuming a Gaussian error distribution with a standard deviation of $\sigma_{D} \approx \delta_{M_{K}} \times 0.2 \times D$ $\times$ ln(10).  From the perturbed sample, in each field, we draw the same number of particles as in our data sample, according to the probability distribution of the ARGOS distances in each field.  Thus, the distances of the mock stars are not matching the initial positions in the simulation, which in turn affects the velocity distribution. In addition, we have assigned a constant velocity error to the mock particles in each field, corresponding to the median uncertainty on each of the 3 velocity components in the data. To summarize, the mock sample has realistic velocity and distance errors, the same number of particles as the data, and the same distance distribution. 

We compute the log-likelihood values for the mock sample on a grid of 180 $\alpha$ values between $-90^{\circ}$ and $+90^{\circ}$, following the same kinematic modelling procedure described in Sec. 4.1. The sum of the ln(L) curves in the 4 central fields is shown in Fig. \ref{logLfinal}, right panel, where $\alpha_{0} = 35^{\circ}$ is the bar angle set for the mock catalog and $\alpha_{\mathrm{min}}  = 35 \pm 3^{\circ}$ is the retrieved bar viewing angle. We repeated this test for numerous random samples and different $\alpha_{0}$ and, despite the small number of mock particles of around $N \sim 400$ per field-of-view (matching the RC sample), the mock catalogue's viewing angle $\alpha_{0}$ can be recovered. The best fit value was found fitting a parabola over an $\alpha$ interval in which ln(L) varies by 9$\sigma$, or  $\Delta \mathrm{ln}(L) = 40.5$ (dotted curve). We have also tested different intervals to check that a $9\sigma$ interval is adequate: an interval of $8-10\sigma$ would produce an identical result and intervals within $5-12\sigma$ would produce a variation smaller than $\pm$3$^{\circ}$, however, the exact input value could be obtained for a $8-10\sigma$ interval. This led us to adopt a   $9\sigma$ interval for the data (left panel of the figure).

\subsubsection{S10 model snapshots}
\begin{figure}
\hspace{-0.5cm}
\includegraphics[width=45.mm]{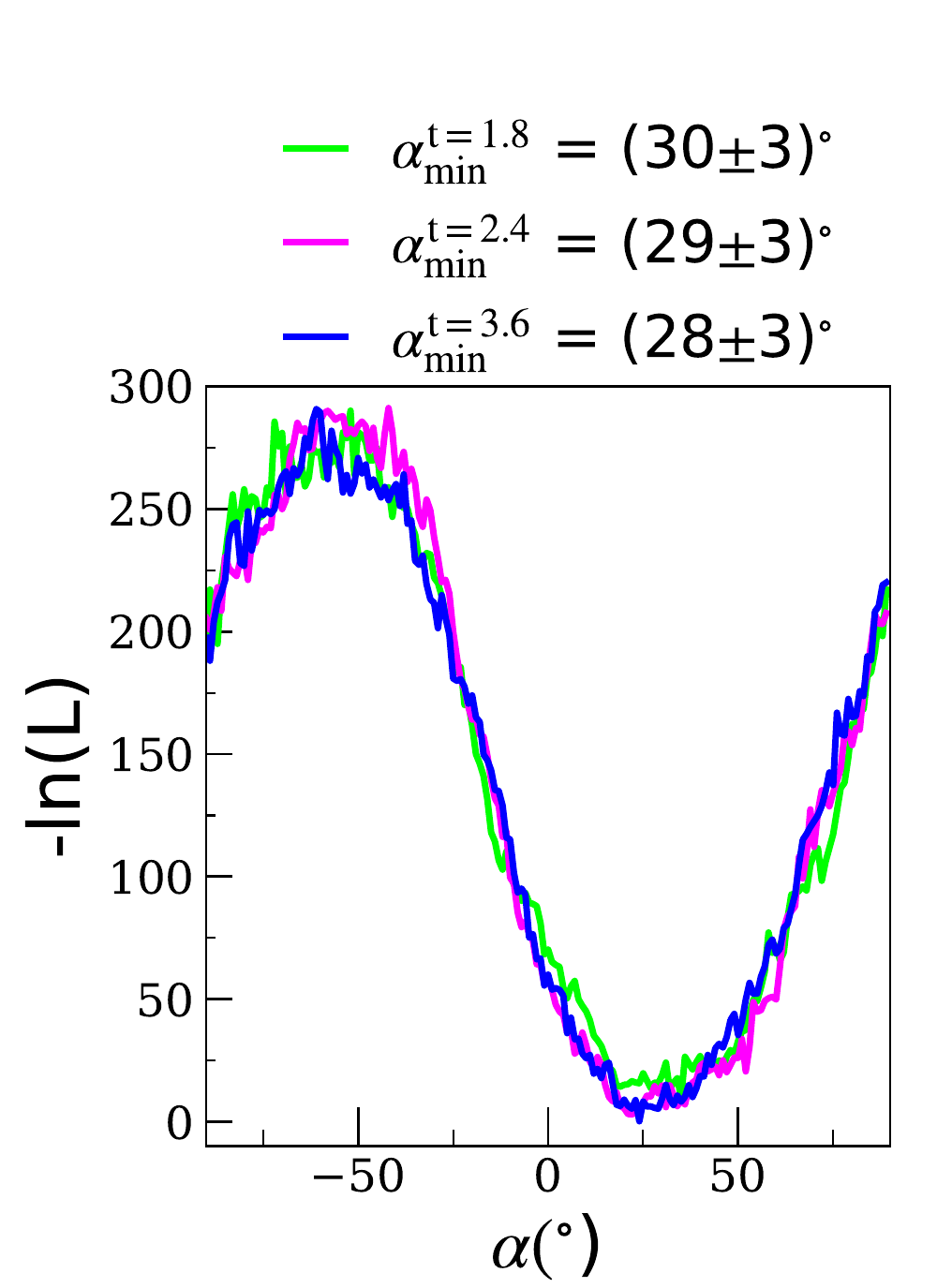} 
\hspace{-0.22cm}
\includegraphics[width=44mm]{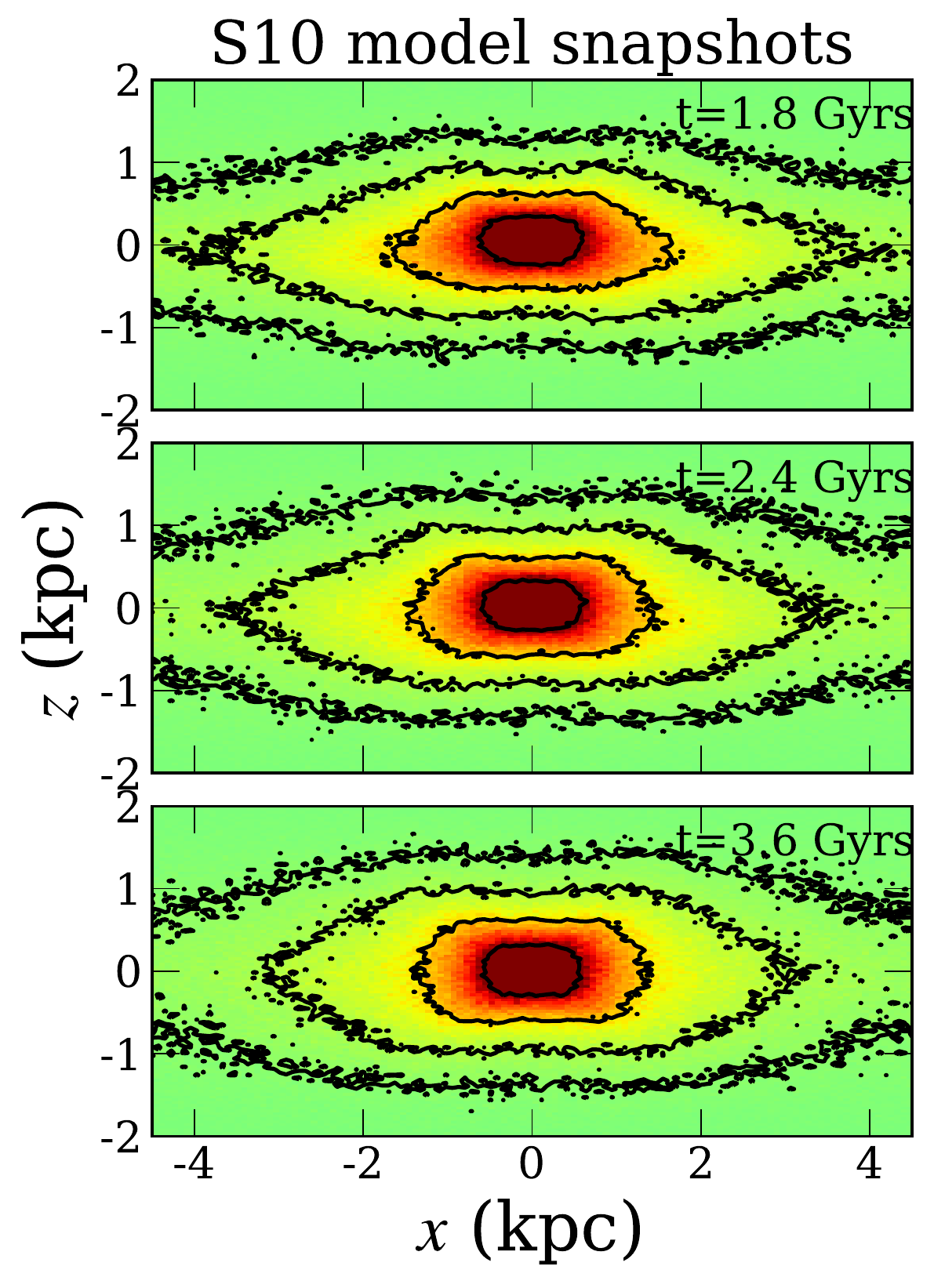} 
\caption{\textit{Left panel:} Same as in the left panel of Fig. \ref{logLfinal} but here the model was constructed using three early snapshots of the S10 model, at 1.8 Gyrs, 2.4 Gyrs and 3.6 Gyrs. Although the models have different density distributions, the angle $\alpha_{0}$ could be recovered successfully. \textit{Right panel:} The density distribution of the S10 model projected on the $x$-$z$ plane at 1.8 Gyrs (top), 2.4 Gyrs (middle) and 3.6 Gyrs (bottom). }
\label{logLallt}
\end{figure}

The S10 model used in this work is a specific instance of a simulated galaxy and while it has been successful at matching observations in the bulge region,  it is likely not a perfect match to the MW due to its simple nature. To check whether small differences in the model density distribution will affect our determination of the bar angle $\alpha$ from the $Gaia$-ARGOS stellar kinematics, we test our kinematic modelling using three snapshots of the S10 model at early times, 1.8, 2.4 and 3.6 Gyrs. Their density distribution projected onto the $x$-$z$ plane is shown in Fig. \ref{logLallt}, right panel. At 1.8 and 2.4 Gyrs the bulge density distribution is not completely symmetric with respect to the Galactic Plane, but at later times (see t = 3.6 Gyrs), as the buckling instability gradually saturates, it becomes increasingly symmetric. MW observations suggest that our bulge is relatively symmetric with respect to the plane, therefore later snapshots may provide a better description of the bulge. We repeat the kinematic modelling procedure using the three earlier snapshots and find that the best estimates of $\alpha$ (Fig. \ref{logLallt}, left panel) are consistent with the value $\alpha_{\mathrm{min}} = 29^{\circ}$ we found using the canonical S10 model at t = 4.8 Gyrs (Fig. \ref{logLfinal}, left panel). The test suggests that the result of our fitting method will not be affected by small changes in the density distribution of the model, likely because of the large distance and $v_{l}$ errors that were implemented in the model to mimick the data.
\begin{figure*}
\includegraphics[width=180mm]{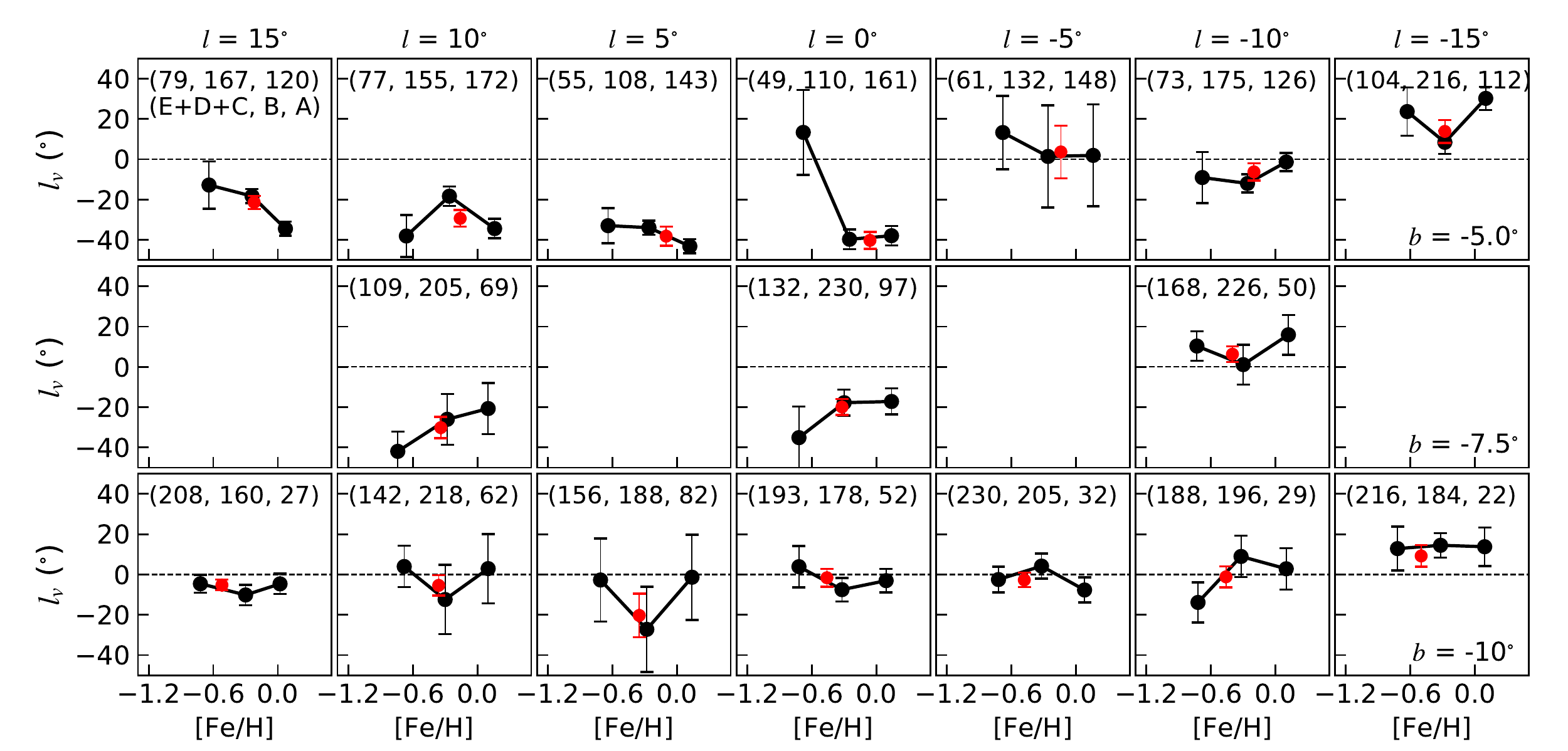}
\caption{Vertex deviation variation with metallicity. For each pointing, the number of stars in each metallicity components is given in the upper left corner. Because of the  bulge metallicity gradient with latitude, there are more metal rich stars close to the Galactic Plane than at lower latitudes. The vertex deviation and the median metallicity for all stars in a given pointing is marked with red.}
\label{met}
\end{figure*} 
\subsubsection{Distance systematic offsets}
A distance offset could be introduced if, for example, the intrinsic magnitude $M_{K}$ of the RC was under/over-estimated. The presence of a younger RC population would also introduce a systematic offset: the RC of a 5 Gyrs population is $-0.1$ mag brighter than that of a 10 Gyrs population, the bulge age commonly assumed.  \citet{Simion2017} estimated that such a population would be predominantely located within $|b|<4^{\circ}$, therefore it should not have a large contribution to the ARGOS fields. \\
The ARGOS collaboration assume $M_{K} = -1.61$ mag for the RC. As a test, we have updated the ARGOS distances assuming two other values, $M_{K} = -1.51$ and $M_{K} = -1.81$. We have then applied the fitting method described in Section 4.1 to the updated datasets. The best fit viewing angles we retrieved are $\alpha = (31\pm 3)^{\circ}$ and $\alpha = (24\pm 3)^{\circ}$ respectively. We note that the typical uncertainty expected for the RC stars $\sigma_{M_{K}} \sim 0.22$ is taken into account in the fitting procedure. \\
We are aware that the parallaxes reported by $Gaia$ DR2 have a systematic offset of $\sim-0.03$ mas \citep{Lind18} which can potentially translate into proper motion systematic offsets, via the cross-terms. Such offsets could affect our computations of $v_{l}$ and $v_{b}$, and finally of $l_{v}$. Performing Monte-Carlo resampling we computed the effect of the parallax offset on the velocities and we found that they change by less than 5 km/s; in fact, the majority of $v_{l}$ (65\%) and $v_{b}$ (80\%) have only changed by 1 km/s. The effect on $l_{v}$ is smaller than $0.1^{\circ}$.
\subsubsection{RGB contamination}
We have also tested how a 30\% contamination rate from the RGB stars would affect our results. In each field of view, we draw new $M_{K}$ values for 30\% of the targets between -3 and +0.5 mag. from an exponential distribution which is commonly used to model the RGB, using the parameters computed by \citet{Simion2017}. We ran the fitting procedure on the new dataset where 70\% of the sample remained unaltered. While the best fit $\alpha$ values in the individual fields vary by $\pm2^{\circ}$ compared to the values reported in Fig. \ref{vertex_alpha}, the final result remains $\alpha = (29 \pm 3)^{\circ}$.
\section{Vertex deviation  as a function of metallicity }
\label{metallicity}
In this section, we investigate the metallicity-kinematics correlation in all ARGOS fields. 

We split the sample in each pointing into three metallicity populations, following the definitions in \citet{Ness2013abundances}: the high metallicity component A, intermediate metallicity component B, and the low metallicity component C. The metal poor components D and E have very few numbers of stars (see their weights in table 3, \citealt{Ness2013abundances}) and we classify them as component C according to the decision boundaries determined using the parameters in tables 2 and 3 in \citet{Ness2013abundances}. As it is apparent from figures 11 and 12 in \citet{Ness2013abundances}, the decision boundaries and the weights of each metallicity component vary with latitude, as is expected due to the bulge metallicity gradient (e.g. \citealt{Gonzalez2013}). In Fig. \ref{met}, we show the vertex measurements in each field as a function of metallicity for all stars in the field (red) and for the 3 sub-populations (black). The $l_{v}$ uncertaintities shown in Fig. \ref{met}, are the standard deviation of 500 $l_{v}$ evaluations on as many bootstrap resampling trials. In the inset we specify the number of stars in each component, with the metal-poor population C+E+D in the left. 

The field at (0$^{\circ}$, -5$^{\circ}$) is the closest to the GC and here $l_{v}$ is most sensitive to changes of $\alpha$ (see Fig. \ref{vertex_alpha}). From Fig. \ref{met}, we notice that it also displays the strongest $l_{v}$ trend with metallicity, in very good agreement with earlier studies and simulations in Baade's window at  $l, b \sim$ (1$^{\circ}$, -4$^{\circ}$) \citep{Soto2007, Babusiaux16review, Debattista2019}: metal rich stars with [Fe/H] $>$ -0.5 have a much higher vertex deviation ($l_{v} \sim$ -40$^{\circ}$) than metal poor stars [Fe/H] $<$ -0.5 ($l_{v} \sim$ 15$^{\circ}$). 

The fields at $b = -10^{\circ}$ are consistent with $l_{v} \sim 0^{\circ}$ at all metallicities as expected (see the last row of Fig. \ref{vertex_alpha}) while for the remaining fields we do not observe obvious trends. The stars at $l>0^{\circ}$ (left panels in Fig. \ref{met}) will likely belong to the near end of the bar, while the stars at $l<0^{\circ}$ (right panels in Fig. \ref{met}), will either be located in the far end of the bar or in the foreground disc population, due to the ARGOS sampling which preferentially targets closer stars. The vertex deviation of an axisymmetric system such as the disc, should be $l_{v} \sim 0^{\circ}$, close to the values we compute in several metallicity bins at $l<0^{\circ}$. 
\section{Conclusion}
\label{conclusion}
We have compiled a sample of $\sim$ 7000 bulge RC stars with 6D phase-space information and metallicity from ARGOS and $Gaia$ DR2. The sample is large enough to allow for a comprehensive study of the bulge kinematics at intermediate latitudes:
\begin{itemize}
\item We mapped the ARGOS-$Gaia$ DR2 velocity distributions in Fig. \ref{vels} and bulge velocity ellipsoids in Fig. \ref{velsgrid}; 
\item  For a specific instance of a simulated galaxy (the S10 model) we have built three maps of the bulge vertex deviation for three bar viewing angles $\alpha = +30^{\circ}, 0^{\circ} , -30^{\circ}$ (Fig. \ref{vertex}). The significant differences between the three maps prove that the bulge morphology has a direct influence on the bulge kinematics; 
\item We have used the S10 model to show that $l_{v}$ varies with $\alpha$, and $l_{v} = 0^{\circ}$ when either of the bulge axes is aligned with the Sun-GC line-of-sight (Fig. \ref{vertex_alpha});
\item We evaluated the probability of our dataset in each individual ARGOS field for a set of kinematic models on a grid of 180 $\alpha$ values (Fig. \ref{logL}), and retrieved the most likely bar angle $\alpha_{\mathrm{min}} = (29 \pm 3)^{\circ}$ via a maximum-likelihood method based on the S10 model.
\end{itemize}
 Our result, derived from kinematic data alone, is an independent measurement of the MW bar viewing angle. While this result is in agreement with $\alpha$ values obtained from star counts studies using millions of stars, we caution the readers that our fitting method is model dependent. In particular, more complex bar models should be used for modelling the kinematics at lower latitudes.  

In the four central fields centred on $(l,b)=(0^{\circ}, -5^{\circ}), (5^{\circ}, -5^{\circ}), (-5^{\circ},5^{\circ}), (0^{\circ},-7.5^{\circ})$, the tilt of the S10 model velocity ellipsoids $l_{v}$ is very sensitive to the bar viewing angle $\alpha$ and it takes values between $l_{v} \sim +45^{\circ}$ and $l_{v} \sim -45^{\circ}$ (Fig. \ref{vertex_alpha}). These four  fields were therefore chosen to derive $\alpha_{\mathrm{min}}$ as the adjacent fields have smaller $l_{v}$ variations with $\alpha$ and do not have as much constraining power. Finally, distance-resampling was a key ingredient of our kinematic modelling. In the top panels of  Figures \ref{vertex} and \ref{vertex_south}, we show the $l_{v}$ map in the S10 model before and after distance-resampling respectively and found that only the latter can match the ARGOS-$Gaia$ DR2 data (Fig. \ref{vertex_south}).  Using the S10 model we have also showed that incompleteness could affect the constraining power of the data: the amplitude of the $l_{v}$ variation with $\alpha$ is larger for a complete sample, as evidenced by the non-resampled S10 model (black curve, Fig. \ref{vertex_alpha}) compared to a resampled model (green).  Future studies using more complete data should be able to determine $\alpha$ also in the outer bulge fields.

The variation of the vertex deviation with metallicity is consistent with previous studies in the Baade's window, with metal-rich stars having a higher vertex deviation than metal-poor stars. With a more complete sample it will be possible to draw firmer conclusions about the vertex deviation variation with metallicity in the remaining fields; however, at low latitudes where the bar does not affect the kinematics, the values are all consistent with $l_{v} \sim 0^{\circ}$ as expected for an axi-symmetric system. 
The Blanco DECam Bulge Survey \citep{BDBS2020} is a photometric survey that can provide metallicities for millions of bulge stars for which $Gaia$ proper motions are readily available. But before we can reproduce the results in Fig. \ref{metallicity} with an increased data set, more radial velocity measurements in the bulge are needed.

\section*{Data availability}
The 2MASS-$Gaia$ DR2 data underlying this article are available in the $\mathrm{gaiadr2.tmass\_best\_neighbour}$ folder, at \url{https://gea.esac.esa.int/archive/}. The ARGOS data were provided by the ARGOS collaboration. Data can be shared on request to the corresponding author with permission of the ARGOS collaboration.

\section*{Acknowledgements}
ITS thanks the referee for the constructive comments that helped improve the manuscript and is grateful for the valuable discussions with V. Belokurov, J. Fernandez-Trincado, A. Robin, Z. Yuan, G. Iorio and V. Debattista. ITS acknowledges support from the PIFI Grant n. 2018PM0050 and LAMOST.  The research presented here is partially supported by the National Key R\&D Program of China under grant no. 2018YFA0404501, by the National Natural Science Foundation of China under grant nos. 11773052, 11333003, 11761131016, and by a China-Chile joint grant from CASSACA. J.S. acknowledges support from an {\it Newton Advanced Fellowship} awarded by the Royal Society and the Newton Fund. SK is partially supported by NSF grants AST-1813881, AST-1909584 and Heising-Simons foundation grant 2018-1030.



\bibliographystyle{mn2e}
\bibliography{bibl}  
\label{lastpage}
\end{document}